\documentclass{vldb}

\usepackage{subcaption}
\usepackage{amsmath, array}
\usepackage{etoolbox}
\usepackage[noend]{algcompatible}
\usepackage{algorithm}
\usepackage{multirow}
\usepackage{makecell}
\usepackage{hyperref}

\usepackage{caption}
\usepackage{booktabs} % For formal tables
\usepackage{subcaption}

\usepackage{graphicx}
\usepackage{balance}  % for  \balance command ON LAST PAGE  (only there!)
\newcolumntype{L}[1]{>{\raggedright\let\newline\\\arraybackslash\hspace{0pt}}m{#1}}
\newcolumntype{C}[1]{>{\centering\let\newline\\\arraybackslash\hspace{0pt}}m{#1}}
\newcolumntype{R}[1]{>{\raggedleft\let\newline\\\arraybackslash\hspace{0pt}}m{#1}}

\def\TSeek{T_{seek}}
\def\TSeqW{T_{seq, W}}
\def\TSeqR{T_{seq, R}}

\def\techReport{}

\vldbTitle{Bridging the Gap Between Theory and Practice on Insertion-Intensive Database}
\vldbAuthors{Sepanta Zeighami, Raymond Chi-Wing Wong}
\vldbDOI{https://doi.org/10.14778/xxxxxxx.xxxxxxx}
\vldbVolume{XX}
\vldbNumber{xxx}
\vldbYear{2020}

\begin{document}
\title{Bridging the Gap Between Theory and Practice on Insertion-Intensive Database}

\numberofauthors{2} %  in this sample file, there are a *total*
% of EIGHT authors. SIX appear on the 'first-page' (for formatting
% reasons) and the remaining two appear in the \additionalauthors section.

\author{
% You can go ahead and credit any number of authors here,
% e.g. one 'row of three' or two rows (consisting of one row of three
% and a second row of one, two or three).
%
% The command \alignauthor (no curly braces needed) should
% precede each author name, affiliation/snail-mail address and
% e-mail address. Additionally, tag each line of
% affiliation/address with \affaddr, and tag the
% e-mail address with \email.
%
% 1st. author
\alignauthor
Sepanta Zeighami\\
       \affaddr{USC}\\
       \email{zeighami@usc.edu}
% 2nd. author
\alignauthor
Raymond Chi-Wing Wong\\
       \affaddr{HKUST}\\
       \email{raywong@cse.ust.hk}
}

\maketitle
\begin{sloppy}

\begin{abstract}
    With the prevalence of online platforms, today, data is being generated and accessed by users at a very high rate. Besides, applications such as stock trading or high frequency trading require guaranteed low delays for performing an operation on a database. It is consequential to design databases that guarantee data insertion and query at a consistently high rate without introducing any long delay during insertion. In this paper, we propose Nested B-trees (NB-trees), an index that can achieve a consistently high insertion rate on large volumes of data, while providing asymptotically optimal query performance that is very efficient in practice. Nested B-trees support insertions at rates higher than LSM-trees, the state-of-the-art index for insertion-intensive workloads, while avoiding their long insertion delays and improving on their query performance. They approach the query performance of B-trees when complemented with Bloom filters. In our experiments, NB-trees had worst-case delays up to 1000 smaller than LevelDB, RocksDB and bLSM, commonly used LSM-tree data-stores, could perform queries more than 4 times faster than LevelDB and 1.5 times faster than bLSM and RocksDB, while also outperforming them in terms of average insertion rate. 
\end{abstract}

\vspace{-5pt}
\section{Introduction}\label{sec:into}

Due to the rapid growth of the data in a variety of applications
such as banking/trading systems \cite{M18}, social media \cite{F17a} and user logs \cite{I17, D17},
 massive data comes in at a rapid rate and it is very important for a database system to handle \emph{both} fast insertion and fast query.
Consider Facebook with more than 41,000 posts \cite{KY17} and YouTube with more than 60,000 videos watched per second on average \cite{DTU14}. The data is generated in a rapid rate and is accessed by other users at the same time. 
Consider Nasdaq Exchange where 
an average of about 70,000 shares  
are traded per second \cite{M18}.
This stock exchange platform requires a database system to guarantee insertion performance at rates higher than 70,000 insertions per second. 
Meanwhile, an insertion delay in an order of milliseconds is unacceptable in many trading scenarios such as in high-frequency trading, a large component of the market \cite{B18} where stocks are traded by milliseconds \cite{N18}. Besides, the current stock price has to be accessed in a short time for the next sell/buy of this stock.

%The database containing this data needs to support fast queries, as the data is shown to other users and is also used in providing recommendations to users. The data generation rate becomes much larger if users' click streams or mouse locations are recorded when visiting the websites.

\subsection{Requirement}

In this paper, we study to design an index which achieves the following 5 requirements.
\begin{enumerate}
    \item \textbf{Short Average Insertion Time Requirement:} The index could handle a lot of insertions within a short period of time. 
    \item \textbf{Short Maximum Insertion Time Requirement:}
    The index could handle each \emph{individual} insertion within a short time.
    \item \textbf{Short Average Query Time Requirement:}
    The index could return the answers of a lot of queries within a short period of time.
    \item \textbf{Short Maximum Query Time Requirement:}
    The index could return the answer of each \emph{individual} query within a very short time.    
    \item \textbf{Theoretical Performance Guarantee Requirement:}
    The index could have theoretical performance guarantee on both the
    insertion performance and the query performance. 
\end{enumerate}

(1) Short Average Insertion Time Requirement is needed due to the rapid data growth nowadays.
(2) Short Maximum Insertion Time Requirement is a \emph{stricter} requirement. It requires that each individual insertion has to be completed within a short period of time but the former requirement requires that the index could handle a collective set of insertions within a period of time, allowing some \emph{individual} insertions to be completed with a \emph{longer} delay.
(3) Short Average Query Time Requirement is needed due to the rapid data access in some applications. (4) Short Maximum Query Time Requirement is needed since it requires that each \emph{individual} query could be answered in a short time.
(5) Theoretical Performance Guarantee Requirement is needed so that we know how good/bad an index is. Based on the first 4 requirements, we are interested in the time complexities of the following
\begin{enumerate}
\item[(a)] Amortized insertion time (Requirement~1)
\item[(b)] Worst-case insertion time (Requirement~2)
\item[(c)] Average query time (Requirement~3)
\item[(d)] Worst-case query time (Requirement~4)
\end{enumerate}
(a) The amortized insertion time of an insertion is the total time the index needs to handle a batch of insertions divided by the total number of insertions handled by the index. 
(b) The worst-case insertion time of an insertion is the greatest insertion time of an insertion. 
(c) The average query time is the query time of a query on expectation.
(d) The worst-case query time is the greatest query time of a query.

Similar to many recent studies \cite{DI19, DI18, DAM+17, CPK+18, BFJ+17, TY17}, we focus on when the data is stored in external memory (e.g., HDD or SSD). Data storage in main memory is more expensive than HDDs or SSDs. As pointed out in \cite{DAM+17} and discussed in \cite{LFA+11}, main memory costs 2 orders of magnitude more than disk in terms of price per bit. Moreover, main memory consumes about 4 times more power per bit than disk \cite{THS10}. Thus, designing high performance external memory indices that provide guarantees for real world applications can significantly reduce the cost of operations for many systems. On Amazon Web Services, any machine with more than 100GB of main memory costs at least US\$1 per hour but a machine with 15.25GB of main memory and 475GB SSD costs US\$0.156 \cite{A19} (Linux machines, US East (Ohio) region). An SSD with 480GB capacity costs US\$55 \cite{A19b} while a 128GB DDR3L RAM module costs about US\$393 \cite{A19c}.     

\subsection{Insufficiency of Existing Indices}
Existing indices do not satisfy the above requirements simultaneously. 
There are two major branches of indices related to our goal:
(1) LSM-tree-like indices \cite{OCG+96, SR12, LHY+10} and (2) B-tree-like indices \cite{BM72, BF03, JDO99} . 

Consider the first branch.
%There are two representative LSM-tree-like indices, namely
%\emph{LSM-trees} \cite{OCG+96, SR12, WXS+15, DFF17, LAK16, DI18, DI19} and \emph{bLSM} \cite{SR12}.
In recent years, LSM-trees \cite{OCG+96, SR12, WXS+15, DFF17, LAK16, DI18, DI19} have attracted a lot of attention and are used as the standard index for insertion-intensive workloads in systems such as LevelDB\cite{G17}, BigTable \cite{CDG08}, HBase \cite{ABC+12}, RocksDB \cite{F17c} (by Google and Facebook \cite{CDG08, ABC+12, F17c}), Cassandra \cite{LM10} and Walnut \cite{CDM+12}. LSM-trees buffer insertions in memory and merge them with on-disk components in bulk, creating sorted-runs on disk. Although LSM-trees satisfy the Short Average Insertion Time requirement, they do not satisfy Short Maximum Insertion Time requirement, Short Average/Maximum Query Time Requirement and Theoretical Performance Guarantee Requirement. This is because LSM-tree's worst-case insertion time is linear in data size \cite{TY17, LHY+10} and their worst-case query time  is suboptimal \cite{LHY+10}. In fact, in our experiments, although RocksDB \cite{F17c}, the industry standard and common research baseline \cite{DI19, DI18, DAM+17}, took an order of microseconds per insertion on average, it had worst-case insertion time of 453 seconds. Such a worst-case insertion time is utterly unacceptable for any application that requires reliability.  

There are two major techniques to improve the performance 
of LSM-trees in the literature. The first technique is \emph{Bloom filters}. 
They can improve average query time of LSM-trees \cite{SR12}, but their worst-case query time remains suboptimal.
Thus, the LSM-trees with Bloom filters still do not satisfy Short Maximum Query Time Requirement.
One representative is 
bLSM \cite{SR12}, a variant of LSM-tree that uses Bloom filters at each level. It also limits the number of LSM-tree levels. Setting the number of LSM-tree levels to a maximum allows for asymptotically optimal query time, but violates Short Average Insertion Time Requirement since the amortized insertion time becomes asymptotically larger than LSM-trees with an unrestricted number of levels. This is because the ratio of the size between LSM-tree components becomes unbounded, causing merge operations to read and rewrite a larger portion of the data. Furthermore, \cite{SR12} provides methods to improve the worst-case query time of LSM-tree by a constant factor, but the worst-case insertion time remains linear to data size. Thus, they still do not satisfy Short Maximum Insertion Time Requirement.
The second technique is \emph{fractional cascading}. It improves the worst-case query time of LSM-trees \cite{LHY+10}, but their average query time remains high.
Thus, LSM-trees with fractional cascading still do not satisfy Short Average Query Time Requirement.
One representative is \cite{LHY+10} that adds an extra pointer to each component of the LSM-tree, pointing to its next component. This pointer allows for reading one disk page per LSM-tree level. This was not compared in the experimental studies of LSM-trees \cite{SR12, DFF17, DI18, DI19} due to its high average query time.
Fractional Cascading and Bloom filters are incompatible \cite{SR12} and cannot be used together.

Consider the second branch. Traditional 
B-trees \cite{BM72} and B$^+$-trees \cite{SKS+97}
are among the most commonly used indices for good query performance. They provide optimal query performance and thus satisfy the Short Average/Maximum Query Time Requirement. They do not satisfy Short Average and Maximum Insertion Time Requirements, because they perform no buffering and perform at least one disk access for every insertion, which is very time-consuming. 

Later, a write optimized variant of B-trees called
B$^\epsilon$-trees (also known as B-trees with Buffer) \cite{BF03}  were proposed, that reserves a portion of each node for a buffer. New data is inserted into the buffer of the root and moved down the levels of the tree whenever the buffer becomes full. However, this method, although faster than B-trees, does not satisfy Short Average/Maximum Insertion Time Requirement. This is because B-tree nodes get scattered across the storage devices and moving this \textit{small} buffer frequently down from a node requires accessing its children which is time consuming.

\subsection{Our Index: NB-Tree}

Motivated by the above, in this paper, we propose an index called the Nested B-tree (NB-tree) which satisfies the 5 requirements simultaneously. That is, NB-trees 
give short average/maximum insertion time which is multiple factors smaller than B-trees and is similar to LSM-trees. They provide worst-case insertion time logarithmic to data size (unlike LSM-tree's linear worst-case insertion time) and multiple factors smaller than B-trees which satisfies the Short Maximum Insertion Time Requirement. They use Bloom filters to provide low average query time, while their structure allows for asymptotically optimal worst-case query time, satisfying Short Maximum Query Time Requirement. 
%Furthermore, NB-trees' support for Bloom filters and asymptotically optimal worst-case query time implies that good practical performance and asymptotically optimal theoretical performance are no longer mutually exclusive. 
This, together with their logarithmic, yet better-than-B-trees, worst-case insertion time shows that NB-trees satisfy the Theoretical Performance Guarantee Requirement.

\begin{figure}
    \centering
    \begin{minipage}[t]{0.49\columnwidth}
        \includegraphics[width=\linewidth]{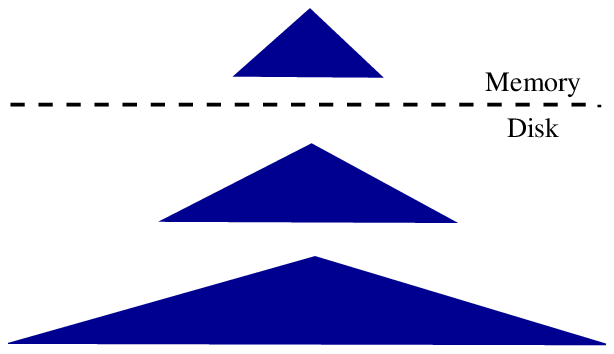}
        \subcaption{LSM-Tree}
        \label{fig:lsm-tree}
    \end{minipage}
    \hfill
    \begin{minipage}[t]{0.49\columnwidth}
        \includegraphics[width=\linewidth]{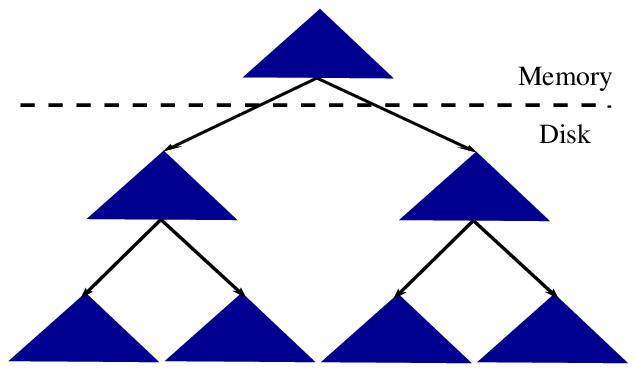}
        \subcaption{Nested B-Tree}
        \label{fig:nb-tree}
    \end{minipage}
    \vspace{-8pt}
    \caption{Nested B-trees break down each level into constant sized B$^+$-trees and establish a connection between the keys in different levels.\vspace{-0.7cm}}
    \label{fig:lsmvsNB}
\end{figure}

Fig. \ref{fig:lsmvsNB} shows the structure of an NB-tree compared with an LSM-tree. Intuitively, a Nested B-tree is a B-tree in which each node contains a B$^+$-tree. NB-trees can be seen as imposing a B-tree structure across the levels of an LSM-tree and breaking down each level into constant-sized B$^+$-trees. By imposing a B-tree structure, NB-trees establish a relationship between the keys in different components and provide an asymptotically optimal query cost, which nears the query performance of B-trees when complemented with Bloom filters. This design is based on the observation that different levels need to be connected to avoid suboptimal worst-case query time, which is lacking in the structure of LSM-trees. Although this is also the intuition behind the design of LSM-tree with fractional cascading \cite{LHY+10}, \cite{LHY+10} fails to design an index compatible with Bloom filters or with logarithmic worst-case insertion time. 

Furthermore, the B-tree structure ensures that keys in each node only overlap with the keys in its children. This limits the impact of merge operations across levels, causing the merge operations to have the same cost on all the levels, and is used to provide a logarithmic worst-case insertion cost. In essence, the connection created between different levels allows us to bound the cost during the merge operation and provide a per-insertion account of the total insertion cost. Such a worst-case analysis is missing in most of the LSM-tree literature \cite{OCG+96, DI19, DI18, DAM+17} where the focus has been on amortized analysis and the few papers that have focused on worst-case performance provide an algorithm with worst-case that is linear to data size \cite{TY17, LHY+10}.   

Finally, by keeping the nodes as large constant-size B$^+$-trees, NB-trees, similar to LSM-trees, perform mainly sequential I/O operations during insertions which minimizes seek time and allows them to perform insertions better than B-trees and their variants.

%\textbf{Contributions.}

\subsection{Contributions and Roadmap}

\begin{itemize}
\vspace{-7pt}
    \item We propose Nested B-Tree (NB-Tree), a novel data structure that satisfies all the 5 requirements mentioned for indices on large volumes of data. This is the first indexing structure satisfying all these 5 requirements in the literature. 
    %\item Bridges the gap between the worst-case and amortized insertion time of LSM-trees;
\vspace{-4pt}
    \item NB-Tree is the first fast-insertion index with asymptotically-optimal query time. To the best of our knowledge, there is no existing index which could achieve this result.
    %\item Provides asymptotically optimal worst-case query time while also supporting Bloom filters;
    %\item Introduces a new method for processing data in batches using a nested structure.
\vspace{-4pt}
    \item In our experiments, NB-Tree's worst-case insertion time was more than 1000 times smaller than LevelDB, RocksDB and bLSM, the three popular LSM-tree databases. %, while on average insertions were performed at least 1.5 times faster. 
    They achieved average query time almost the same as B$^+$-trees, while performing insertions at least 10 times faster than them on average. 
\end{itemize}

\begin{table*}  
	\centering
	\begin{tabular}{|C{5cm}|C{2.1cm}|C{2.1cm}|C{2.1cm}|C{2.1cm}|C{2.2cm}|}\hline
	       Data Structure &   Amortized Insertion Time &  Worst-Case Insertion Time  & Average Query Time   & Worst-Case Query Time  & Asymptotically Optimal Query Time \\\hline
	B-Tree \cite{BM72} and B$^+$-Tree  &   Bad &   Medium   & Good & Good & Yes \\\hline
	B-Tree with Buffer \cite{BF03} &   Medium &   Medium   & Medium & Medium & Yes \\\hline
	LSM-Tree (no BF, no FC) \cite{OCG+96}  &   Good &   Bad   & Bad & Bad & No \\\hline 
	LSM-Tree (BF, no FC) LevelDB \cite{G17}, RocksDB \cite{F17}, Monkey \cite{DAM+17}   &   Good &   Bad   & Medium    & bad  & No\\\hline
	LSM-Tree (no BF, FC) \cite{LHY+10}  &  Good   &   Bad &   Bad   & Medium  & Yes\\\hline
	NB-Tree (BF) [this paper]  &   Good &   Good   & Good   & Medium  & Yes\\\hline
    \end{tabular}
    \vspace{-15pt}
    \caption{Comparing NB-trees, LSM-trees and B-tree variants (BF: with Bloom filters, FC: with fractional cascading)\vspace{-0.6cm}}
    \label{tab:requirementResults}
\end{table*}

\vspace{-4pt}
\textbf{Summary of Results.} The performance improvement of NB-trees compared with LSM-tree variants and B-tree is summarized in Table \ref{tab:requirementResults}. NB-trees outmatch LSM-trees on worst-case insertion and query time as well as average query time, and perform insertions faster than B-trees while providing similar average query performance. A more in-depth analysis of the related work is provided in Sec.~\ref{sec:relatedWork}.

\textbf{Organization.} The rest of this paper is organized as follows. Section \ref{sec:term} discusses the terminology used and the problem addressed in this paper. Section \ref{sec:NBTreeDesign} provides the design of the Nested B-tree data structure and Section \ref{sec:imp} discusses more details on the implementation and analysis of the data structure. Section \ref{sec:advanced} discusses a more advanced version of NB-tree that achieves a logarithmic worst-case insertion time and uses Bloom filters. Section \ref{sec:exp} provides our experimental results. Section \ref{sec:relatedWork} discusses the relevant literature and Section \ref{sec:conclusion} provides the conclusion of the paper.

\section{Terminology and Setting}\label{sec:term}
\textbf{Problem Setting.} Key-value pairs are to be stored in an index that supports insertions, queries, deletions and updates. The index is to be stored on an HDD or SSD and the term \textit{disk} is used to broadly refer to the secondary storage device. The data is written or read from disk in pages of size $\mathcal{B}$ bytes. The index can use up to $M$ pages of main memory. Transferring a page from disk to main memory (or vice versa) incurs two costs, a \textit{seek} time, $T_{seek}$, and a \textit{sequential} read, $T_{seq, R}$ or write, $T_{seq, W}$, time. Seek time is the time difference between the starting time of
the read/write request 
and the starting time of the data transfer.
%latency from the time the read or write request is sent to the disk until the time the data transfer starts and 
\textit{sequential} read/write time is the time taken to transfer the data from the disk to the main memory.

Sequential access time is determined by a device's bandwidth while seek time depends on its internal mechanisms: on HDDs the movement of the disk arm and platter, on SSDs the limitation of its electrical circuits. Sequential time is proportional to size of the data transferred but seek time depends on how the data is stored, i.e., whether it is stored on contiguous blocks. It is important to account for seek time in our analysis as, per page, it can take much longer than sequential access time. For instance, an HDD (7200rpm and 300MB/s bandwidth) based on the measurements in \cite{S17} has a seek time of 8.5 milliseconds and transfer rate of 125 MB/s. Reading a 4KB disk page incurs seek time of $8.5\times 10^{-3}$ seconds, but reading it sequentially takes $\frac{4KB}{125MB/s} \approx 3\times 10^{-5}$ seconds (283 times smaller than the seek time). 

For the ease of discussion and as is the industry standard for common key-values stores such as RocksDB \cite{F17} and LevelDB \cite{G17}, we consider the keys to be unique. Duplicate keys can be handled similar to B-trees \cite{SKS+97} by using an extra bucket or a uniquifier attribute as discussed in \cite{SKS+97}.

\textbf{Performance Metrics.} When analyzing an index we assume its performance is dominated by disk I/O operations. For an operation on an index (e.g, an insertion or query on the index) we use the term \textit{cost} only when referring to the \textit{number of pages} accessed during the operation. We use the term \textit{time} when referring to the \textit{actual time} taken, measured in seconds, during the operation. The time is dominated by disk I/O operations, and is composed of the sequential and seek time for all disk accesses performed during the operation. Because a \textit{time} measure takes into account seek operations, it is a more realistic measure of the real-life performance of an index compared with the \textit{cost} measures.  

We use the following metrics for evaluation of indices. \textit{Worst-case insertion time} is the time, measured in seconds, it takes to insert an item into the index in the worst case. Moreover, given a set $X$ of $n$ keys to be inserted to an index, \textit{amortized insertion time} of a key in $X$ with respect to $X$ is the worst-case total time of inserting all the keys of $X$ divided by $n$. \textit{Worst-case query time} is the time an index takes to answer a query in the worst-case. \textit{Average query time} is the expected value of the random variable denoting the query time of a random query key (\textit{average} time is defined over one operation and is the expected time the operation takes while \textit{amortized} time is defined over a set of operations, and is the average time an operation takes in the worst-case). The metrics are defined in terms of \textit{time}, but their definition in terms of \textit{cost} is analogous.

\textbf{Problem Definition.} Our goal is designing an index that satisfies the Short Average/Maximum Insertion Time, Short Average/Maximum Query Time and Theoretical Performance Guarantee Requirements. %These requirements translate to an index that performs well, theoretically and in practice, based on the performance metrics discussed above. 

\vspace{-7pt}
\section{Design of Nested B-tree}\label{sec:NBTreeDesign}
%Next, we introduce Nested B-tree (or NB-tree), an index designed to satisfy the 5 requirements mentioned before. 
An NB-tree is a B-tree whose nodes contain B$^+$-trees. NB-trees insertion, deletion and update operations differ from that of B-trees, but the data is organized in the nodes of an NB-tree in a way that the properties of B-trees (in addition to other properties described later) are preserved. This allows for query and worst-case insertion time that grows only logarithmically in data size. Moreover, insertion, deletion and update operations are first buffered in memory and batched together which reduces the number of page accesses and the number of seek operations performed, improving significantly on amortized and worst-case insertion times compared with B-trees. These properties, complemented with Bloom filters, allow NB-trees to support insertions and queries at high rates without any delays during insertion. 

Next, we describe a basic version of NB-trees. We provide the final version in Section \ref{sec:advanced}. We use Fig. \ref{fig:NBTreeExampleInsert} for illustration.

\begin{figure*}[t!]
%\begin{minipage}[t]{1\textwidth}
	\centering
		\includegraphics[width=1\textwidth]{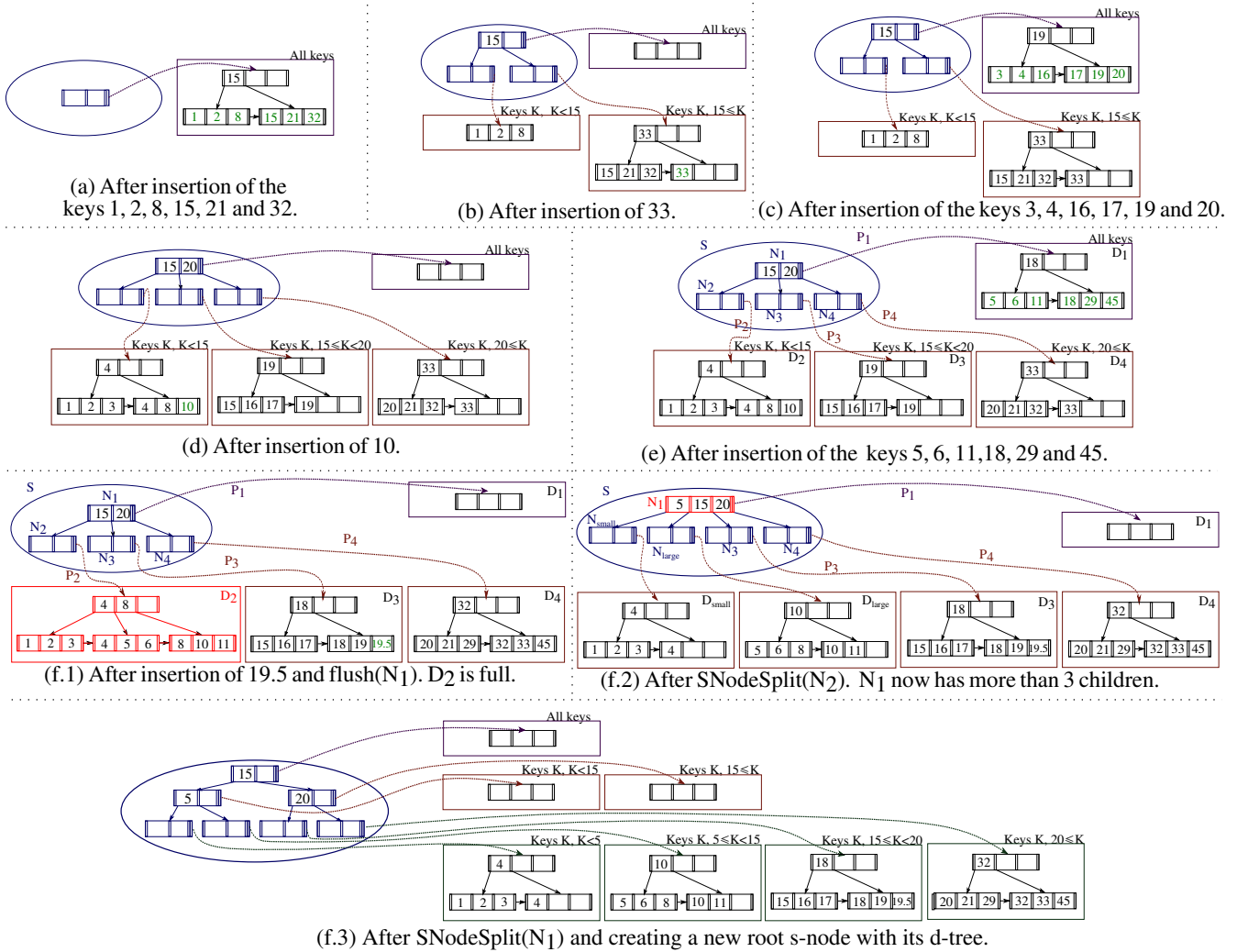}
        \vspace{-17pt}
		\caption{Insertions in an NB-Tree with parameters $\sigma = 6$ key-value pairs, $f = 3$, $B = 4$.}
		\label{fig:NBTreeExampleInsert}
%\end{minipage}
%\end{figure*}
\vspace{-15pt}
\end{figure*}

\subsection{Overview, Definitions and Properties}\label{sec:design:overview}
An NB-tree is defined as a collection of several tree structures, $\{D_1, ..., D_k, S\}$ for an integer $k$. 

$D_1$ to $D_k$ are B$^+$-trees that each store part of the data (i.e., key-value pairs) and are called \textit{data trees} or \textit{d-trees} for short. Any key-value pair inserted into the index is stored in one of the d-trees, and the key-value pairs are moved between the d-trees throughout the life of an NB-tree. In Fig. \ref{fig:NBTreeExampleInsert} (e), $D_1$, $D_2$, $D_3$ and $D_4$ show four different data trees. They are all B$^+$-trees, i.e., at the leaf level each key is written next to its corresponding value (not shown in the figure). For ease of discussion we refer to the nodes of a data tree as \textit{data nodes} or \textit{d-nodes} and to the keys in a d-node as \textit{data keys} or \textit{d-keys}

$S$ is a tree structure similar to a B-tree. $S$ is used to establish a relationship between the keys in the d-trees, and impose a structure on the d-trees. Thus, $S$ is called a \textit{structural tree} or an \textit{s-tree} for short.  A structural tree is exactly a B-tree with some modifications discussed later. In Fig. \ref{fig:NBTreeExampleInsert} (e), the eclipse labelled $S$ shows a structural tree. Similar to a B-tree, an s-tree contains several nodes. For ease of discussion we refer to the nodes of a structural tree as \textit{structural nodes} or \textit{s-nodes} and to the keys in an s-node as \textit{structural keys} or \textit{s-keys}. 

An s-tree differs from a B-tree in the following ways. (1) An s-tree does not store any key-value pairs. It only contains keys and pointers. Keys in an s-tree are not associated with a value. For this reason we call it a structural tree (it only specifies a structure). (2) Each s-node, $N$, contains an extra pointer to the root d-node of a d-tree (which is a B$^+$-tree). We call this d-tree, $N$'s d-tree (each d-tree is pointed to by exactly one s-node). The pointer in an s-node pointing to the root of its d-tree will be referred to as its \textit{d-tree pointer}. In Fig. \ref{fig:NBTreeExampleInsert} (e), pointers $P_1$, $P_2$, $P_3$ and $P_4$ are d-tree pointers for s-nodes $N_1$, $N_2$, $N_3$ and $N_4$. (3) Leaf s-nodes only contain a d-tree pointer, and no keys or values.  This is because an s-tree does not contain any data in its s-nodes. Since leaf s-nodes don't have any children, they do not contain any pointers or keys. In Fig. \ref{fig:NBTreeExampleInsert} (e), leaf s-nodes $N_2$, $N_3$ and $N_4$ do not contain any keys and only contain a d-tree pointer.

Specifically, non-leaf s-nodes in an s-tree are of the format $\langle P_{d-tree}, P_1, K_1, P_2, K_2, ..., P_r, K_r, P_{r+1}\rangle$ for an s-node with $r+1$ children. $P_i$ for all $i$ are pointers to the s-nodes in the next level of the s-tree, $K_i$ are the corresponding s-keys and $P_{d-tree}$ is a pointer to the d-tree of the s-node. s-keys are sorted in an s-node. For an s-key, $K$, in the s-node pointed to by $P_i$, $1<i<r+1$, it is true that $K_{i-1}\leq K< K_i$, for $i=1$, $K< K_i$ and for $i=r+1$, $K_{i-1}\leq K$. The only differences between a non-leaf s-node and a non-leaf B-tree node is that (1) an s-node has an extra pointer $P_{d-tree}$ to the d-tree of the s-node and (2) s-keys are not associated with any value in the s-node. Moreover, a leaf s-node is of the format $\langle P_{d-tree}\rangle$, i.e., it only contains a d-tree pointer.

\vspace{-5pt}
\subsubsection{Properties} %NB-trees satisfy the following properties. 

\textbf{Structural Properties}. The following properties are the structural properties of NB-trees. 

\textit{S-tree Fanout.} Each non-leaf s-node has at most $f$ children and each non-leaf and non-root s-node has at least $\lceil\frac{f}{2}\rceil$ children. We call the parameter $f$ the \textit{s-tree fanout}. In Fig. \ref{fig:NBTreeExampleInsert}, $f$ is set to 3. Each s-node has at most 3 children, and non-leaf and non-root s-nodes must have at least 2 children.

\textit{D-tree Fanout.} Each non-leaf d-node has at most $B$ children and each non-leaf and non-root d-node has at least $\lceil\frac{B}{2}\rceil$ children. We call the parameter $B$ the \textit{d-tree fanout}. In Fig. \ref{fig:NBTreeExampleInsert}, $B$ is set to be 4. Each d-node has at most 4 children, and non-leaf and non-root s-nodes must have at least 2 children.

\textit{D-tree Size.} For a parameter $\sigma$, each d-tree is at most of size $\sigma$. D-trees of leaf but not root s-nodes are at least of size  $\lceil\frac{\sigma}{2}\rceil$. $\sigma$ can be specified by the number of bytes used by the d-tree or the number of key-value pairs in the d-tree. The analysis in the paper uses the latter for ease of notation, while the former is used in experiments as its easier to specify in practice. Unless stated otherwise, $\sigma$ refers to the number of key-value pairs in a d-tree (i.e., number of d-keys in the leaf level). In Fig. \ref{fig:NBTreeExampleInsert}, $\sigma$ is set to 6. Each d-tree contains up to 6 keys in their leaves, and d-tree of leaf but not root s-nodes contain at least 3 d-keys in their leaves.

\textbf{Cross-s-node Linkage Property.} This property of NB-trees establishes the relationship between the s-keys in an s-node and the d-keys and s-keys of the s-node's children. Consider an s-node, $N$ with $r$ s-keys. The s-node is of the form $\langle P_{d-tree}, P_1, K_1, P_2, K_2, ..., P_r, K_r, P_{r+1}\rangle$, where the s-keys are in a sorted order (i.e., for each $i \in [1, r), K_i< K_{i+1}$). For each $i$, $1<i< r+1$, consider the child s-node, $C_i$ pointed to by $P_i$. For an s-key in $C_i$ or a d-key in $C_i$'s d-tree, $K_{C_i}$, it holds that $K_{i-1} \leq K_{C_i} < K_i$ . For $i = 1$, it holds that $K_{C_i}< K_i$ and for $i=r+1$, $K_{C_i}\geq K_{i-1}$. In Fig. \ref{fig:NBTreeExampleInsert} (e), the d-keys in $D_2$ are less than 15, the d-keys in $D_3$ are at least 15 but less than 20 and the d-keys in $D_4$ are at least 20. In Fig. \ref{fig:NBTreeExampleInsert}, d-trees are labelled with a possible range that is derived based on this property. The possible range of d-keys for s-nodes in the same level is non-overlapping and covers the entire key space. 

%\textbf{Self-balancing Property.} An s-tree, similar to B-trees, is self-balancing. We use a splitting mechanism similar to B-trees to achieve this.

%\textbf{Partially Sorted  Property.} Consider the keys in d-trees of the s-nodes on the same level of the s-tree. These keys are fully sorted. For instance in Fig. \ref{fig:NBTreeExampleInsert} (e), because of the cross-s-node linkage property, keys in $D_2$ are less than $D_3$ and all of them are less than the keys in $D_4$. However, the keys among different levels are not fully sorted. For instance, the relationship between the keys in $D_1$ and $D_2$ is not known. 

\vspace{-6pt}
\subsection{Operations}
\vspace{-8pt}

\subsubsection{Insertions}\label{sec:design:insertion}
The insertion of a key-value pair $(K, V)$ in an NB-tree starts by inserting the pair in the d-tree of the root s-node and recursively moving the pair down the tree to ensure that the properties mentioned in Section \ref{sec:design:overview} are satisfied. We refer to a d-tree as \textit{full} if it has more than $\sigma$ key-value pairs. In this section, we provide a conceptual discussion on how insertions are performed, and how they are implemented in practice is discussed in Section \ref{sec:implementation}.  

Intuitively, the d-tree of each s-node can be seen as a storage space for the s-node. The key-value pairs are stored in the d-tree of each s-node. When d-tree of an s-node $N$ becomes \textit{full}, the pairs are distributed down to the d-tree of the children of $N$ based on the $N$'s s-keys such that the Cross-s-node Linkage Property is satisfied. This continues until the d-tree, $D$ of a leaf s-node, $N$ becomes full, in which case $D$ and $N$ are split into two and the median of d-keys in $D$ (i.e., the d-key, $K$, in $D$ such that half of the d-keys in $D$ are less than $K$) is inserted into the parent, $P$, of $N$. If $P$ now has more than $f$ children, $P$ and it's d-tree are similarly split into two. The splitting may continue until the root of the s-tree, which may result in an increase in the height of the tree. %The splitting process is in fact very similar to that of B-trees, except that (1) the splitting of the leaf s-nodes occurs when the d-tree of a leaf s-node becomes full, and not the s-node itself and (2) the d-tree of an s-node is also split into two when the s-node is split. 
More specifically, insertion works as follows. 

\textbf{Insertion Operation.} A new key-value pair $(K, V)$ is always inserted into the d-tree, $D$, of the root s-node, $N$. We insert $(K, V)$ in $D$ using a B$^+$-tree insertion mechanism. If $D$ has up to $\sigma$ d-keys, the insertion is finished. Otherwise, we need to ensure \textit{d-tree size requirement} is satisfied. For this, we call $HandleFullSNode(N)$ (described later).  

%we distribute d-keys of $D$ to have space for the insertion into $D$ without violating the . We call the $HandleFullSNode(N)$ operation that resolves this issue. If $HandleFullSNode(N)$ returns $NULL$, it means that $D$ have been emptied. In this case, we now insert $(K, V)$ in $D$ using a B$^+$-tree insertion mechanism and the insertion is done.  

%If $HandleFullSNode(N)$ does not return $NULL$, it returns the tuple $(K_M, P_{small}, P_{large})$. This means that moving the d-keys out of $D$ has consequently resulted in $N$ being split into two new s-nodes (we describe when this happens later). In this case, we need to create a new root s-node for the s-tree as the previous root s-node, $N$, is split into two s-node (this is similar to when the root of a B-tree is split into two). Consider two nodes $N_{small}$ and $N_{large}$ that are the result of the split. In this scenario, $HandleFullSNode(N)$ returns the median key, $K_M$, of the keys in $N$ together with pointers $P_{small}$ and $P_{large}$ to $N_{small}$ and $N_{large}$. Then, we create a new s-node $R$ and a new d-tree $D_R$, set $R$'s d-tree pointer to $D_R$ and insert $K_M$, $P_{small}$ and $P_{large}$ in $R$. This new s-node will be the root of the s-tree. Then we insert the key-value pair $(K, V)$ in the d-tree of this new root s-node .  

\textit{Example.} In Fig. \ref{fig:NBTreeExampleInsert} (a), insertion of key-value pairs is done in the d-tree of the root s-node. Fig. \ref{fig:NBTreeExampleInsert} (a) shows the result of inserting keys 1, 2, 8 ,15, 21 and 32. They are all inserted into the d-tree of the root s-node. Now, inserting a new key (e.g., 33) in the d-tree of the root s-node causes the d-tree to become full and $HandleFullSNode$ is called on the root s-node to restore compliance to the \textit{d-tree size requirement}. Fig. \ref{fig:NBTreeExampleInsert}(b) shows the result after calling $HandleFullSNode$ and all the properties discussed in Section \ref{sec:design:overview} are satisfied. 

%Five more d-keys are inserted into the root as shown in Fig. \ref{fig:NBTreeExampleInsert} (e) until the root s-node becomes reaches size $\sigma$. To insert a new key (e.g., 12) now we call the $HandleFullSNode$ operation. This time the operation splits the root into two. So we create a new root and insert the median returned by $HandleFullSNode$ in the root s-node. Now we can insert 12 in the d-tree of this new root (Fig. \ref{fig:NBTreeExampleInsert} (f.3)). Note that when root is also a leaf s-node (Fig. \ref{fig:NBTreeExampleInsert} (a)), $HandleFullSNode$ will split the root in which case we also need to create a new root s-node and insert the median returned by $HandleFullSNode$ into the new root s-node (Fig. \ref{fig:NBTreeExampleInsert} (b)).

\textbf{HandleFullSNode Operation}. $HandleFullSNode(N)$ is called to restore compliance to \textit{d-tree size requirement}  when the size of a d-tree, $D$, of an s-node $N$ surpasses $\sigma$. It acts differently when $N$ is a leaf s-node and when it is not.
%If $N$ is a leaf s-node, it resolves the issue by splitting $D$ and $N$ into two. If $N$ is not a leaf s-node, it resolves the issue by moving the d-keys of $D$ to the d-trees of $N$'s children. This algorithm may result in $N$ being split into two s-nodes, in which case, it returns the median s-key of $N$ (or median d-key if $N$ is a leaf) together with the pointers to two newly created s-nodes that are results of the split. It returns $NULL$ otherwise. There are two possibilities.

\textit{N is a leaf s-node.} $HandleFullSNode(N)$, if $N$ is a leaf s-node, calls $SNodeSplit(N)$. $SNodeSplit(N)$ (detailed later) splits $N$ into two s-node $N_{small}$ and $N_{large}$ and returns the median d-key, $K_M$, of the d-keys in $D$ together with pointers $P_{small}$ and $P_{large}$ to $N_{small}$ and $N_{large}$. Then $HandleFullSNode(N)$ inserts $K_M$, $P_{small}$ and $P_{large}$ into the parent s-node of $N$ and returns (similar to the insertion of the median into a parent node of a B-tree after the node splits). If $N$ is a root s-node, i.e., has no parent, $HandleFullSNode(N)$ creates a new root s-node and then inserts $K_M$, $P_{small}$ and $P_{large}$ into this new root (s-tree's height increases by one). 

\textit{Example.} Consider Fig. \ref{fig:NBTreeExampleInsert} (a). $HandleFullSNode$ splits the d-tree and the s-node into two, one d-tree containing the smaller half and another the larger half of d-keys (seen at the leaf level of Fig. \ref{fig:NBTreeExampleInsert} (b)). $HandleFullSNode$ also creates a new root s-node and inserts the median of d-keys into it. 

\textit{N is not a leaf s-node.} If $N$ is not a leaf s-node, $HandleFullSNode(N)$ first calls $flush(N)$ operation. $flush(N)$ (detailed later) removes the keys-value pairs from $D$ and inserts them into the d-trees of $N$'s children. After that, for any child s-node $C$ of $N$, if $C$'s d-tree is now full, $HandleFullSNode(N)$ calls $HandleFullSNode(C)$ recursively. If d-tree of none of the children is full, $HandleFullSNode(N)$ returns. 

If $N$ has $k$ children, there can be up to $k$ recursive calls. A recursive call $HandleFullSNode(C)$ may result in $C$ being split into two s-nodes, which increases the total number of children of $N$. Therefore, if the number of children of $N$ becomes larger than $f$, $HandleFullSNode(N)$ calls $SNodeSplit(N)$ which splits $N$ into s-nodes $N_{small}$ and $N_{large}$ and returns the median s-key $K_M$ of $N$ together with pointers $P_{small}$ and $P_{large}$ to $N_{small}$ and $N_{large}$. Then $HandleFullSNode(N)$ inserts $K_M$, $P_{small}$ and $P_{large}$ into the parent s-node of $N$ and returns (similar to the insertion of the median into a parent node of a B-tree after the node splits). If $N$ is a root s-node, i.e., has no parent, $HandleFullSNode(N)$ creates a new root s-node and then inserts $K_M$, $P_{small}$ and $P_{large}$ into the new root (s-tree's height increases by one). 

%$N_{small}$ and $N_{large}$ as a result of the recursive call, the recursive call returns the median key, $K_M$, of the keys in $C$ together with pointers $P_{small}$ and $P_{large}$ to $N_{small}$ and $N_{large}$. Then, $K_M$ is inserted into $N$ together with $P_{small}$ and $P_{large}$. The recursive calls can increase the total number of children of $N$ because of the splits. Therefore, if the number of children becomes larger than $f$ then $N$ will itself split into s-nodes $N_{small}$ and $N_{large}$. In this case, the operation will return the median s-key $K_M$ of $N$ with two pointers to $N_{small}$ and $N_{large}$. Otherwise, it returns $NULL$. 

\textit{Example.} Consider Fig. \ref{fig:NBTreeExampleInsert} (e). $HandleFullSNode(N_1)$ first calls $flush(N_1)$ which moves d-keys from d-tree of the root s-node $N_1$ to its children. Fig. \ref{fig:NBTreeExampleInsert} (f.1) shows the result. Consequently $N_2$'s d-tree becomes full (has more than 6 keys). Thus, $HandleFullSNode(N_1)$ calls itself recursively, i.e., $HandleFullSNode(N_2)$. In the recursive call, since $N_2$ is a leaf s-node, $HandleFullSNode(N_2)$ calls $SNodeSplit(N_2)$ which splits $N_2$ into two. It inserts the median d-key into $N_1$. Now $N_1$ has more than two s-keys (Fig. \ref{fig:NBTreeExampleInsert} (f.2)). 

Then $HandleFullSNode(N_1)$, calls $SNodeSplit(N_1)$ which splits $N_1$ into two. It creates a parent for $N_1$ and inserts $N_1$'s median, 15, into $N_1$'s parent. Fig. \ref{fig:NBTreeExampleInsert} (f.3) shows the result.

\textbf{SNodeSplit.} $SNodeSplit(N)$ splits an s-node $N$ and its corresponding d-tree $D$ into two. If $N$ is a leaf s-node, let $K_M$ be the median d-key of $D$. If $N$ is not a leaf s-node, let $K_M$ be the median s-key of $N$. $SNodeSplit(N)$ creates two s-nodes $N_{small}$ and $N_{large}$ with corresponding d-trees $D_{small}$ and $D_{large}$. It inserts all the d-keys in $D$ less than  $K_M$ in $D_{small}$ and the d-keys at least $K_M$ in $D_{large}$. It also inserts all s-keys in $N$ less than $K_M$ in $N_{small}$ and the s-keys at least $K_M$ in $N_{large}$. Let $P_{small}$ be a pointer to $N_{small}$ and $P_{large}$ a pointer to $N_{large}$. The operation returns $(K_M, P_{small}, P_{large})$.

\textit{Example.} See Fig. \ref{fig:NBTreeExampleInsert} (f.1) to (f.2) and Fig. \ref{fig:NBTreeExampleInsert} (f.2) to (f.3). 

\textbf{Flush}. $flush(N)$ is called on a non-leaf s-node, $N$. Intuitively, $flush(N)$ distributes the d-keys in the d-tree of $N$ to the d-tree of its children based on the s-keys of $N$. Let $N$ contain $r$ s-keys and be of the format $\langle P_{d-tree}, P_1, K_1, P-2, K_2, ..., P_r, K_r, P_{r+1}\rangle$. Let $D$ denote the d-tree of $N$, pointed to by $P_{d-tree}$. Furthermore, let $C_i$ be the s-node pointed to by $P_i$ and let $D_i$ be the d-tree of $C_i$. For every key $K$ in $D$, we remove it from $D$ and insert it into $D_i$ if $K_{i-1}\leq K< K_i$. We insert $K$ into $D_1$ if $K< K_1$ and in $D_{r+1}$ if $K\geq K_{r}$.

\textit{Example.} See Fig. \ref{fig:NBTreeExampleInsert} (e) to Fig. \ref{fig:NBTreeExampleInsert} (f.1). 

%\textbf{SNode Insertion.} At several parts of the algorithm, we insert $(K, P_{small}, P_{large})$ in an s-node $N$. This is done similar to insertions in a B-tree node. Consider an s-node with $r$ keys of the format $\langle P_{d-tree}, P_1, K_1, P_2, K_2, ..., P_i, K_i, P_{i+1}, ..., P_r, K_r, P_{r+1}\rangle$. To do this, we find the smallest $i$ such that $K_i>K$. We insert $K$ right before $P_i$, $P_{small}$ right before $K$ and we set $P_i$ to $P_large$. If no such $i$ exists, we set $P_{r +1}$ to $P_{small}$ and write $K$ and $P_{r+1}$ at the end of the s-node.

\begin{figure*}[t!]
%\begin{minipage}[t]{1\textwidth}
    \centering
		\includegraphics[width=0.8\textwidth]{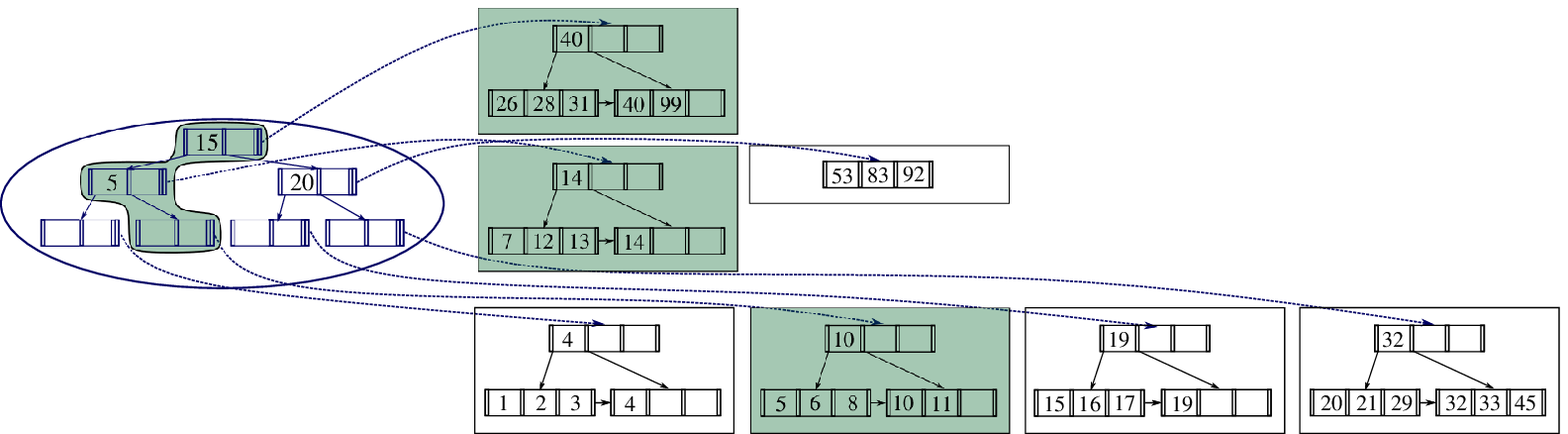}
		\vspace{-7pt}
		\caption{Querying for the key 11 in an NB-Tree. The shaded area shows the part of the tree read during the query.}
		\label{fig:NBTreeExampleQuery}
\vspace{-15pt}
%\end{minipage}
\end{figure*}

\subsubsection{Updates and Deletions} Similar to LSM-trees \cite{OCG+96}, we perform deletion and updates by inserting delta records that indicate the modification into the index. Thus, deletions and updates are treated the same way as insertions and the same analysis applies to them. Note that delta records will be resolved before they reach the leaf level or they can be discarded (if they reach the leaf level, the key they are meant to modify does not exist in the tree). Therefore, delta records do not affect the height of the s-tree and do not affect our analysis of query and insertion performance. The cases when s-nodes become \textit{underfull} as a result of a deletion, can also be handled using a mechanism similar to deletions in B-trees \cite{SKS+97}. We will not provide the details for such scenarios since our focus is on write-intensive workloads, and we assume that deletions are infrequent and such cases can be ignored.

\subsubsection{Queries}\label{sec:design:query}
NB-trees perform queries similar to B-trees. However, in a B-tree, a query traverses the B-tree based on the keys in the nodes. In an NB-tree, a query traverses the s-tree based on the s-keys in the s-nodes. Furthermore, in a B-tree, a query only searches the keys in the nodes visited. However, an NB-tree searches the s-keys in the s-nodes visited and also searches the corresponding d-tree of each s-node visited. Searching a d-tree is exactly a B$^+$-tree search. %A query, $q$, is done by searching the root s-node, $N$ for the key that is to be found. Searching $N$ is done by a typical B$^+$-tree search performed on the d-tree of the s-node. Let the leaf d-node reached by the B$^+$-tree search be $M$. If the key exists in $M$, or if $N$ is a leaf s-node, the query is finished. Otherwise, consider the d-node $M$. If $M$ contains an s-key, $K$, such that $K > q$, then consider the s-key s-pointer, $L$, of the smallest such $K$. If not, consider the d-node s-pointer, $L$, of the d-node $M$ (such a pointer exists because $N$ is a non-leaf s-node). Then, the query is performed by recursively searching the NB-tree rooted at the s-node pointed to by $L$. Range queries can also be performed similarly. 
Fig. \ref{fig:NBTreeExampleQuery} shows the query of key 11 on an NB-tree.%TODO: range queries

\section{Implementation and Analysis}\label{sec:imp}
%We discuss the implementation of an NB-tree and provide its performance analysis. 
To allow for fast performance, similar to LSM-trees, the d-tree corresponding to the root s-node is kept in memory. The rest of the d-trees are stored on disk. 

\subsection{Insertion Implementation}\label{sec:implementation}
Manipulations of the s-tree is straight forward. Here we focus on operations impacting on-disk d-trees. $Insert$ and $HandleFullSNode$ do not make any modifications to the on-disk d-tree themselves. Modifications are done through $flush$ and $SNodeSplit$ operations, so we focus on them. 

To minimize the insertion time, we aim at minimizing the number of seek operations by performing our disk accesses sequentially. To this end, we maintain the following invariants. Firstly, all the d-nodes in a d-tree are written sequentially and can be retrieved by a sequential scan from the first node written. Secondly, the leaf d-nodes are written on disk in a sorted order. Thus, a sequential scan of a d-tree from the first leaf d-node until the last d-node reads all the key-value pair written in the d-tree in an ascending order.

\textbf{Flush(N)}. Assume that $N$ contains $r+1$ children, $C_i$ with respective d-trees $D_i$, $1\leq i\leq r+1$ and $r$ keys $\langle K_1, K_2, ..., K_r\rangle$. $flush$ starts by sequentially scanning $D$ and $D_1$, merge-sorting them together (sequential scan of $D$ and $D_1$ retrieves their keys in a sorted order) and writing the output, $D_1'$ in a new disk location. Note that the two invariants mentioned above now hold for $D_1'$. From $D$, we only merge-sort the d-keys that are less than $K_1$ with $D_1$. We follow the same procedure and in general merge-sort the d-keys, $K$ such that $K_{i-1} \leq K< K_{i}$, from $D$ with $D_i$ for $1<i<r$. For $i=r+1$, we merge-sort the d-keys, $K$ such that $K_{i} \leq K$, from $D$ with $D_{r+1}$ and for $i=1$ d-keys, $K$, such that $K_{i} > K$. Finally, we move down only the first $\sigma$ d-keys from $D$ if it has more. This is to avoid the size of the full d-trees in deeper levels of the tree getting progressively larger as a result of recursive $flush$ calls. Because some of the d-keys may remain in a d-tree, we re-write $D$ starting from the $(\sigma +1)$-th d-key and thus removing the d-keys that were flushed down from $D$. %Note that as a result of a flush operation, a child $C_i$ can now have at most $2s$ keys ($s$ moved from the parent and $s$ from before) before $flush(C_i)$ is called and its size is reduced to $s$ keys.

The cost of $flush$ is  $O(\frac{\sigma f}{B})$. Assuming the main memory has enough space to buffer $\Omega(\sigma)$ key-value pairs (to buffer a constant fraction of the parent's d-tree and the d-tree of one child at a time) which is typically in the order of 100MB, the flush operation performs a constant number of seek operations for merge-sorting $N$ with each child and thus $O(f)$ seek operations in total. The number of seek operations increases proportionately if there is less space available in memory. 

\textbf{SNodeSplit(N)} The $SNodeSplit(N)$ operation only performs disk accesses when dividing a d-tree into two. For this, we sequentially scan a d-tree and sequentially write it as two d-trees. It costs $O(\frac{\sigma}{B})$ page accesses and $O(1)$ number of seek operations under the same conditions as above. This operation preserves the two invariants mentioned above. 

%Note that because of the invariants mentioned before, our writing of d-trees is done sequentially, and no in-place modification is done on the data structure. This allows us to retrieve the data from each d-tree and write the data to the disk with sequential scans, resulting in a low number of seek operations performed as analyzed in Section \ref{sec:analysis}. 

\subsection{Analysis}\label{sec:analysis}
%Here, we analyze the performance of NB-trees based on the implementation discussed above.

\textbf{Correctness.} Induction on the number of insertion operations shows that the cross-s-node linkage and structural properties are preserved using the insertion algorithm. The correctness of the query operation follows from the cross-s-node linkage property, and the correctness of updates and deletions follow from the correctness of insertions. 

\textbf{Insertion Time Complexity.} There are at most $O(\frac{n}{\sigma})$ $HandleFullSNode$ function calls on any level because in the worst case all the keys are moved down to the leaf level and each $flush$ moves $\sigma$ keys. $HandleFullSNode$, excluding the recursive call, requires $O(\frac{f\sigma}{B})$ page accesses for $flush$ and $SNodeSplit$. Each operation can be handled with $O(f)$ seek operations. Since the height of the s-tree is $O(\log_f \frac{n}{\sigma})$, the amortized insertion time is $O(\log_f(\frac{n}{\sigma})\times (\frac{f}{B}\TSeqW + \frac{f}{\sigma}\TSeek))$. Note that we only modify an s-node if its corresponding d-tree is modified. Thus, assuming each s-node fits in a disk page ($f$ is typically much smaller than $B$) s-tree manipulations add at most one page write after writing each d-tree, which does not impact the complexity of the operations. 

For this version of NB-tree, the worst-case insertion time is linear in $n$ because all the s-nodes may be full at the same time. In Section \ref{sec:advanced} we introduce a few modifications that reduces the worst-case insertion time to logarithmic in $n$.

\textbf{Query Time Complexity.} In the worst case, the query will search one s-node in each level of the s-tree. The height of each d-tree is $O(\log_B \sigma)$ and height of the s-tree is $O(\log_f \frac{n}{\sigma})$, thus, the query takes time $O(\log_B \sigma \log_f \frac{n}{\sigma})\times(\TSeek+\TSeqR)$. Observe that the query cost of NB-trees is asymptotically optimal. That is, it is within the constant factor $\log_B \sigma$ of minimum number of pages accesses required to answer a query. Note that in-memory caching, to cache a number of levels of each d-tree can be used to reduce query time by a constant factor, similar to B-trees. 

%Finally, note that, similar to LSM-trees \cite{OCG+96}, because we do not overwrite any parts of the d-trees during the insertions, queries can easily performed while an insertion operation is being executed. We only need to ensure that d-tree pointers in each s-node is changed when a d-tree is fully written. Thus, queries do not need to wait for the insertion to be finished before they are performed. 

\subsection{Parameter Setting}
NB-trees have three parameters, $f$, $\sigma$ and $B$. $B$ is set similar to B-trees so we focus on the other two. $f$ provides a trade-off between insertion cost and query cost while $\sigma$ provides a trade-off between the number of seek operations per insertion and query cost. $\sigma$ depends on how expensive seek operations are, but typically, for fast insertions, it is set to the order of tens or hundreds of mega bytes. Typically, $f$ is set to a number in the order of 10 for write intensive workloads, and its increase affects insertions much more than queries, as the insertion time linearly depends on $f$ but query time's dependence is only logarithmic. Section \ref{exp:small:param} provides an empirical analysis of parameter setting.

\section{Advanced NB-tree}\label{sec:advanced}
We discuss modifications to the NB-tree design to reduce the worst-case insertion time from linear in $n$ to logarithmic in $n$ and how to add Bloom filters to NB-trees to enhance their query performance. %With these modifications, we create the \textit{advanced version} of NB-tree that performs insertions at high insertion rates with very low insertion delay, while answering queries very fast. 
The version provided here is to be considered as the final NB-tree index.%TODO: amoritzed cost twice LSM-trees. We can only do it in one scan

\subsection{Modification}
We make the following changes to the structural properties of NB-tree. For non-leaf s-nodes, we remove the requirement on the maximum size of its d-tree being $\sigma$, and instead put a requirement on the total number of key-value pairs in the d-trees of all sibling s-nodes to be $f(\sigma+1)$ (each s-node can still have at most $f-1$ keys). We also restrict $f$ to be at most a constant fraction of $\sigma$ which is typically true in practice.

\textbf{Single Recursive Call.} All the operations work the same as before, but with one difference. In $HandleFullSNode(N)$, after calling $flush(N)$, if any s-node is oversized, $HandleFullSNode$ will be called recursively on exactly one s-node that has the largest size (i.e. $\arg\max |C|$), instead of performing a recursive call for every full s-node. The rest of the operations work as before, but now there is at most one recursive call during $HandleFullSNode$ operation.

The above insertion procedure remains correct and satisfies the new requirement on the maximum number of key-value pairs in d-trees of non-leaf sibling s-nodes. This is because each level receives $\sigma$ keys and flushes down $\sigma$ keys (see Section \ref{sec:implementation}) if any of the d-trees of sibling s-nodes have more than $\sigma$ keys, and the requirement is already satisfied if none of the siblings has more than $\sigma$ keys. For leaf s-nodes we still perform splits if their size surpasses $\sigma$ keys. Thus, we can observe that the total size of siblings is at most $f\times \sigma$.

\textbf{Lazy Removal.} Recall that during the flush operation, we need to remove the d-keys that were moved from the parent s-node to its children. In Section \ref{sec:implementation} we discussed a method that required rewriting of the parent s-node. Here, we discuss a lazy removal approach that removes this overhead. Consider the scenario when $flush(N)$ is called, assume that $N$'s parent is $P$ and $N$'s d-tree is $D$. Some of the d-keys of $D$ are flushed to the d-tree of children of $N$. At this stage, we create a pointer to the location of the smallest d-key $K$ in $D$ that is not flushed to $N$'s children, that is, all the d-keys in $D$ smaller than $K$ are now present in the d-tree of $N$'s children and need to be removed from $D$. Now instead of removing these d-keys from $D$ at this point, we postpone this removal to when $flush(P)$ is called (i.e., when $N$ is a child s-node during the flush operation). When $flush(P)$ is called, we need to flush the d-keys from $P$'s d-tree to $N$'s d-tree. In doing so, we only merge d-keys in $D$ that are at least equal to $K$ (using the pointer to $K$ we remembered). Because of the sequentiality property, these keys can be retrieved by a sequential scan, and the existence of the keys smaller than $K$ in $N$ does not incur any extra cost for $flush(P)$. After $flush(p)$ is called, an entirely new d-tree is created for $N$ and we discard the previous d-tree now, removing the keys smaller than $K$. This lazy removal does not incur any extra cost for insertions as the d-keys whose removal where postponed will not be read by the insertion algorithm. Moreover, the total size of siblings will be $f(\sigma+1)$ because one s-node can now have at most $\sigma$ more s-keys than was discussed in the above paragraph. 

\textbf{Deamortization.} Although the worst-case insertion time of NB-tree with the changes discussed above is already logarithmic in data size (shown below), we deamortize the insertion procedure by performing $\frac{1}{\sigma}$ fraction of the operations for every new key inserted into the NB-tree, similar to \cite{LHY+10}, to reduce the worst-case insertion time by the factor $\sigma$. 

\textbf{Insertion Time Complexity.} An insertion operation performs at most one $HandleFullSNode$ function call at each level of the s-tree, resulting in at most $O(\log_{f}\frac{n}{\sigma})$ number of $HandleFullSNode$ calls. Each $flush$ and $SNodeSplit$ step take $O(\frac{f\sigma}{B})$ I/O operations. Thus, the total time take for one insertion call is $O(\log_{f}\frac{n}{\sigma}(\frac{f\sigma}{B}\TSeqW + f\TSeek))$. Deamortization reduces the cost by a factor of $\sigma$ and we can achieve the worst-case insertion time $O(\log_{f}\frac{n}{\sigma}(\frac{f}{B}\TSeqW + \frac{f}{\sigma}\TSeek))$. The amortized insertion time in this case is the same as the worst-case insertion time. This shows that NB-trees achieve a good amortized and worst-case insertion time compared with LSM-trees and B-trees, as shown in Table \ref{tab:requirementResults}. %Note that this is the same as the amortized insertion time for M\&S (but as shown above, the worst-case cost is improved from linear in $N$ to logarithmic in N).

%\textit{BM\&S in the Background and Insertion Delay.} Note that when the root s-node becomes full, we can run BM\&S in the background, that empties the root s-node, and continue performing insertions in the emptied root s-node in the foreground. As long as the main memory s-node has space, the insertions will not make any disc accesses and therefore can be performed while the BM\&S operation is working in the background (performing queries  while BM\&S is running in the background is discussed in Section \ref{sec:concurrency}). This can reduce insertion delay significantly.

\textbf{Query Time Complexity.} Maximum size of an s-node is $f(\sigma+1)$ since that is the maximum total size of sibling s-nodes together. Thus, the query cost is now at most $O(\log_B(f\sigma)\times (\log_f(\frac{n}{\sigma})))$ based on an analysis similar to Section \ref{sec:design:query}, but by changing the maximum size of an s-node. $f$ is at most a fraction of $\sigma$ and thus $\log_B(f\sigma)$ is  $O(\log_B(\sigma^2))$ which is $O(\log_B(\sigma))$. Hence, the query cost is $O(\log_B(\sigma)\times (\log_f(\frac{n}{\sigma})))$, which is asymptotically optimal as discussed in Table \ref{tab:requirementResults}. 

\subsection{Bloom Filter}\label{sec:bloom}
We use Bloom filters to enhance the average query cost. A Bloom filter uses $k$ bits per key and $h$ hash functions to decide whether a key exists in a data structure. When searching for a key, if the Bloom filter returns negative, the key definitely does not exist in the data structure. When it returns positive, the key may not exist in the data structure with a probability dependant on $k$ and $h$ (e.g., $k = 8$ and $h=3$ results in a false positive probability of less than 5$\%$). 

We use a Bloom filter for d-tree of each s-node. We need to create/modify the Bloom filters during $flush$ or $SNodeSplit$ operations. For children s-nodes in $flush$ and all the s-nodes in $SNodeSplit$, as we create a new d-tree for the s-nodes, we create a new Bloom filter for this d-tree and delete the old Bloom filter if it exists. For the parent s-node in $flush$, as mentioned above we use lazy removal, that is, the d-tree is kept until the s-node is a child in a $flush$ operation, when a new d-tree is created and the old d-tree discarded. We similarly keep the Bloom filter and create a new one only when the s-node is a child in the $flush$ operation.

To search for a key $q$, we start our search from the root s-node. We check if the Bloom filter for the root indicates that the d-tree of the root can contain $q$ or not. If yes, then we search the root. If it does not contain $q$, then we move down one level according to the pointers and perform the search recursively on the subtree rooted at the node. Overall, in the worst case, we go through all the levels of the s-tree and search the corresponding d-tree, which gives the same worst-case query time as before. However, with high probability, we only search one s-node in total and the cost will be $O(\log_B \sigma)$ with high probability, which is a constant. Thus, NB-trees have a good average query time, as mentioned in Table \ref{tab:requirementResults}.

\begin{figure*}[t]
    \begin{minipage}[t]{.49\textwidth}
    	\includegraphics[width=\textwidth]{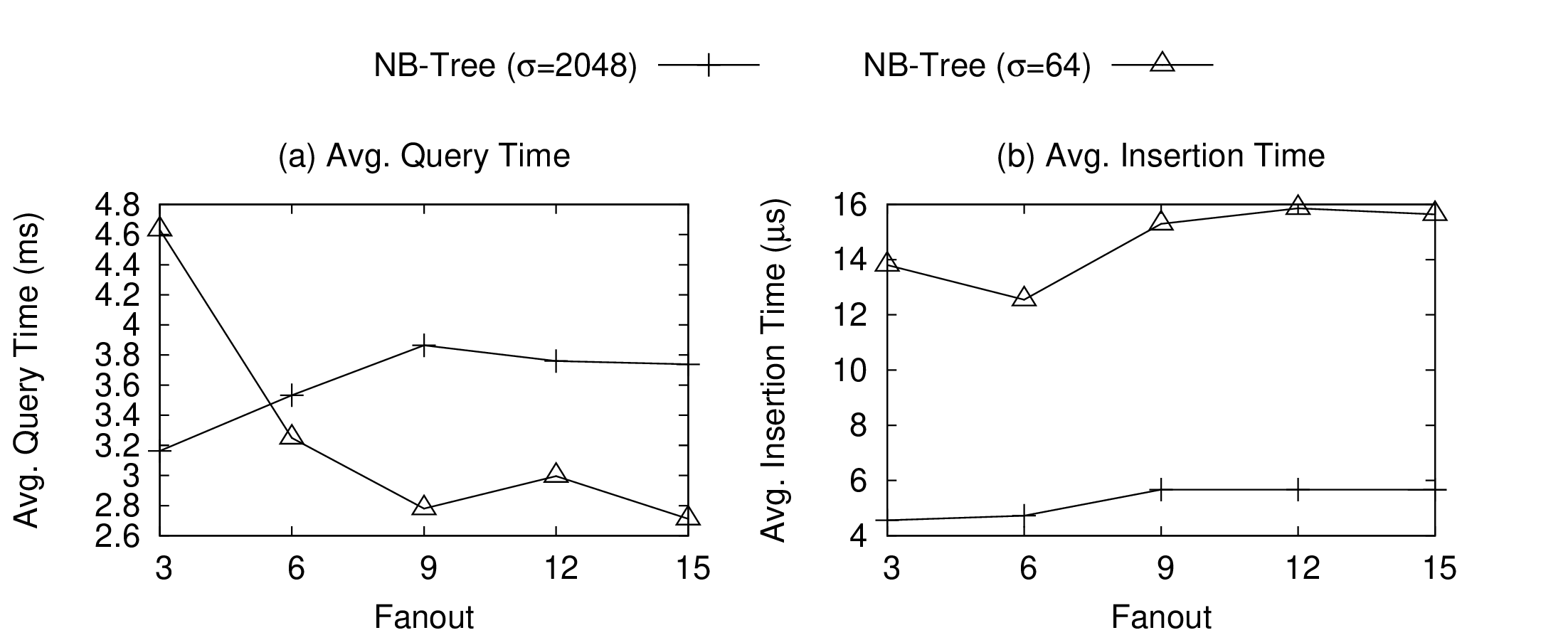}
        \vspace{-0.7cm}
    	\caption{NB-Tree Performance vs Fanout}
    	\label{fig:exp:paramF}
    \end{minipage}
    \begin{minipage}[t]{.49\textwidth}
    	\includegraphics[width=\textwidth]{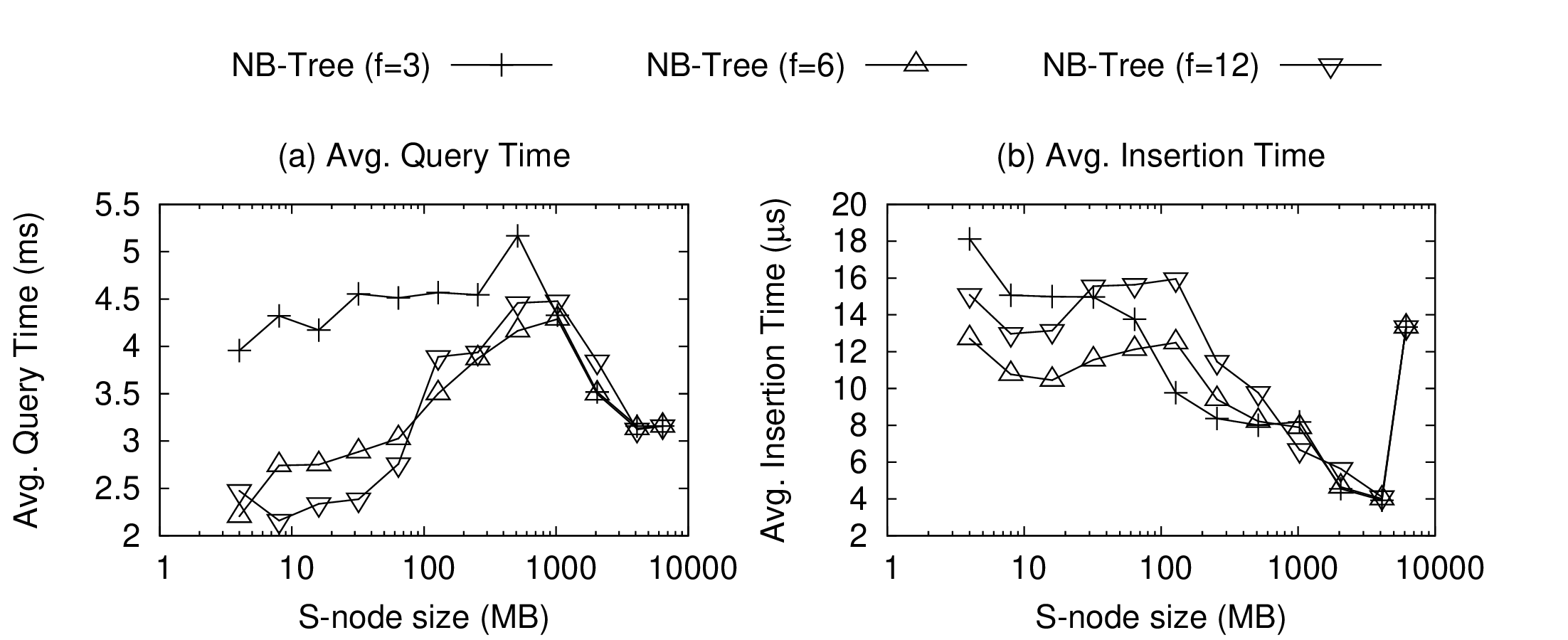}
        \vspace{-0.7cm}
    	\caption{NB-Tree Performance vs. S-Node Size}
    	\label{fig:exp:paramS}
    \end{minipage}
    \vspace{-15pt}
\end{figure*}
\begin{figure*}[t]
    \begin{minipage}[t]{.49\textwidth}
    	\includegraphics[width=\textwidth]{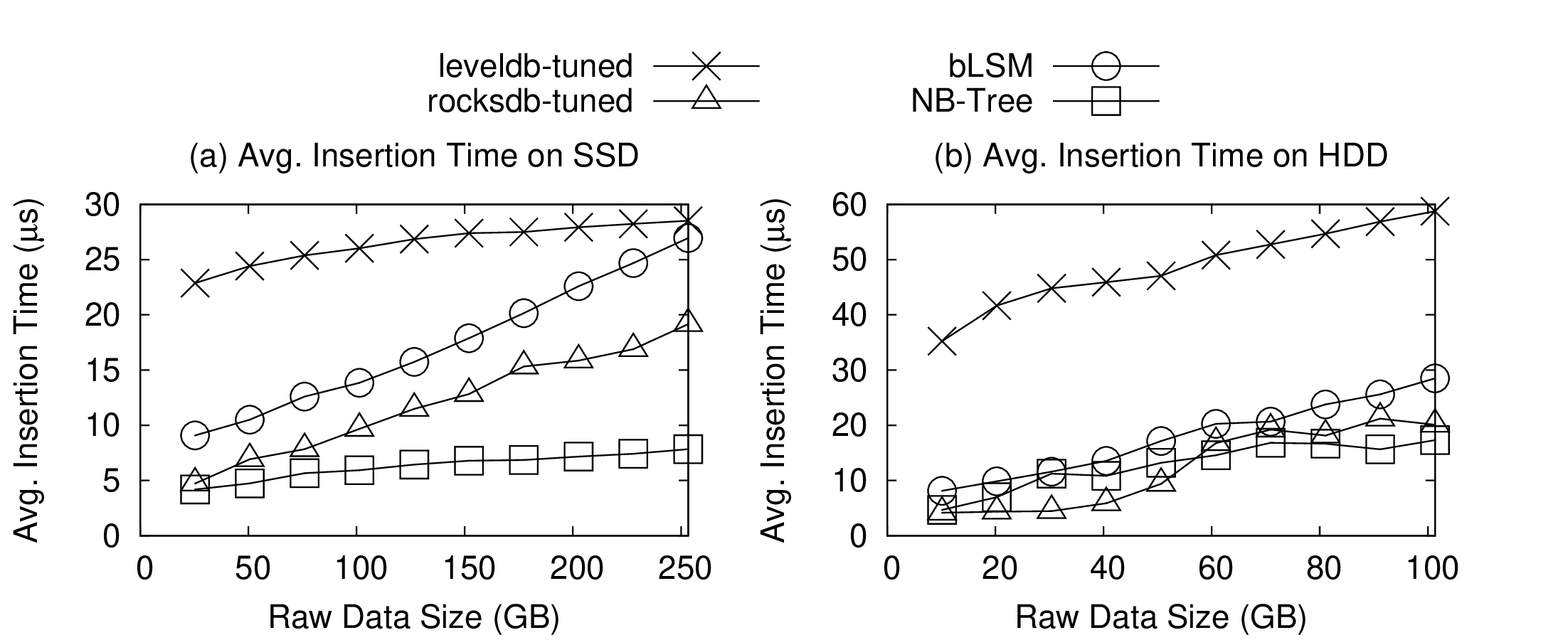}
        \vspace{-0.7cm}
    	\caption{Avg. Insertion Time  vs. Data Size}
    	\label{fig:exp:insertion}
    \end{minipage}
    \begin{minipage}[t]{.49\textwidth}
    	\includegraphics[width=\textwidth]{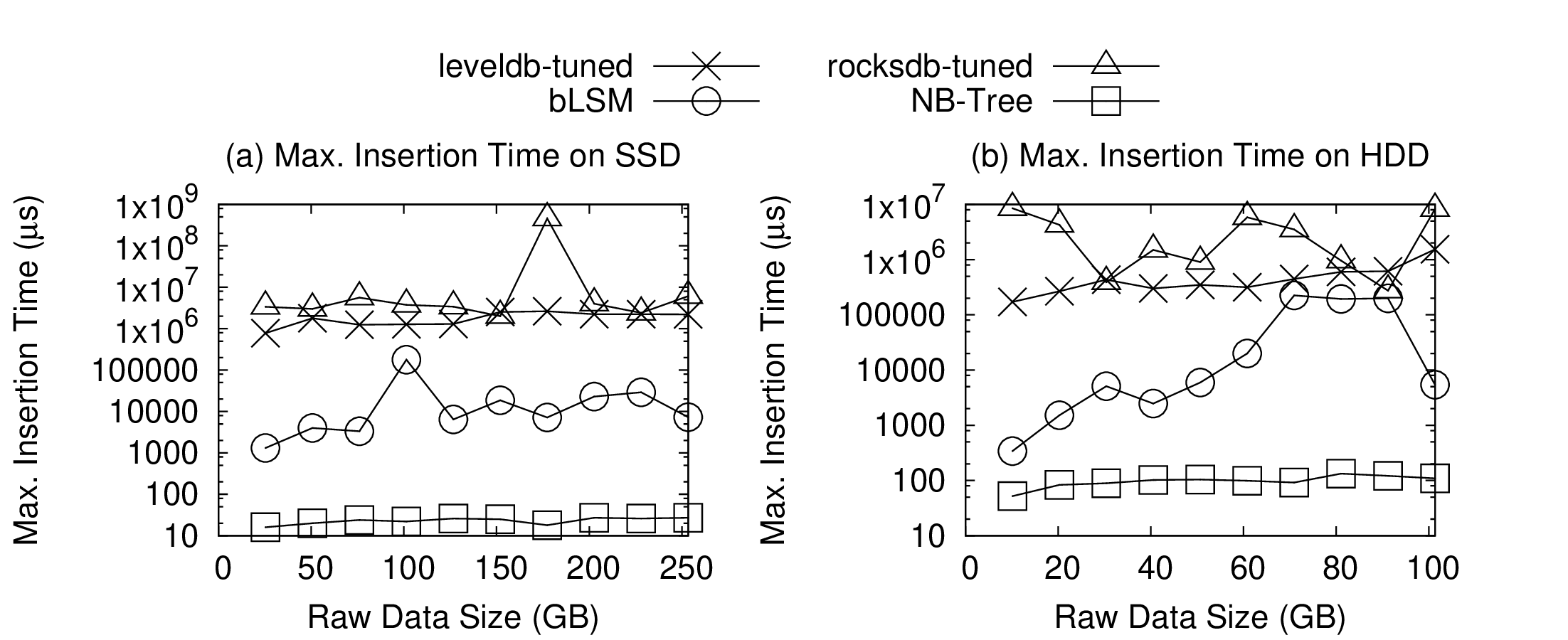}
        \vspace{-0.7cm}
    	\caption{Max. Insertion Time  vs. Data Size}
    	\label{fig:exp:maxinsertion}
    \end{minipage}
    \vspace{-15pt}
\end{figure*}

\begin{figure*}[t]
    \begin{minipage}[t]{.49\textwidth}
        \centering
    	\includegraphics[width=\textwidth]{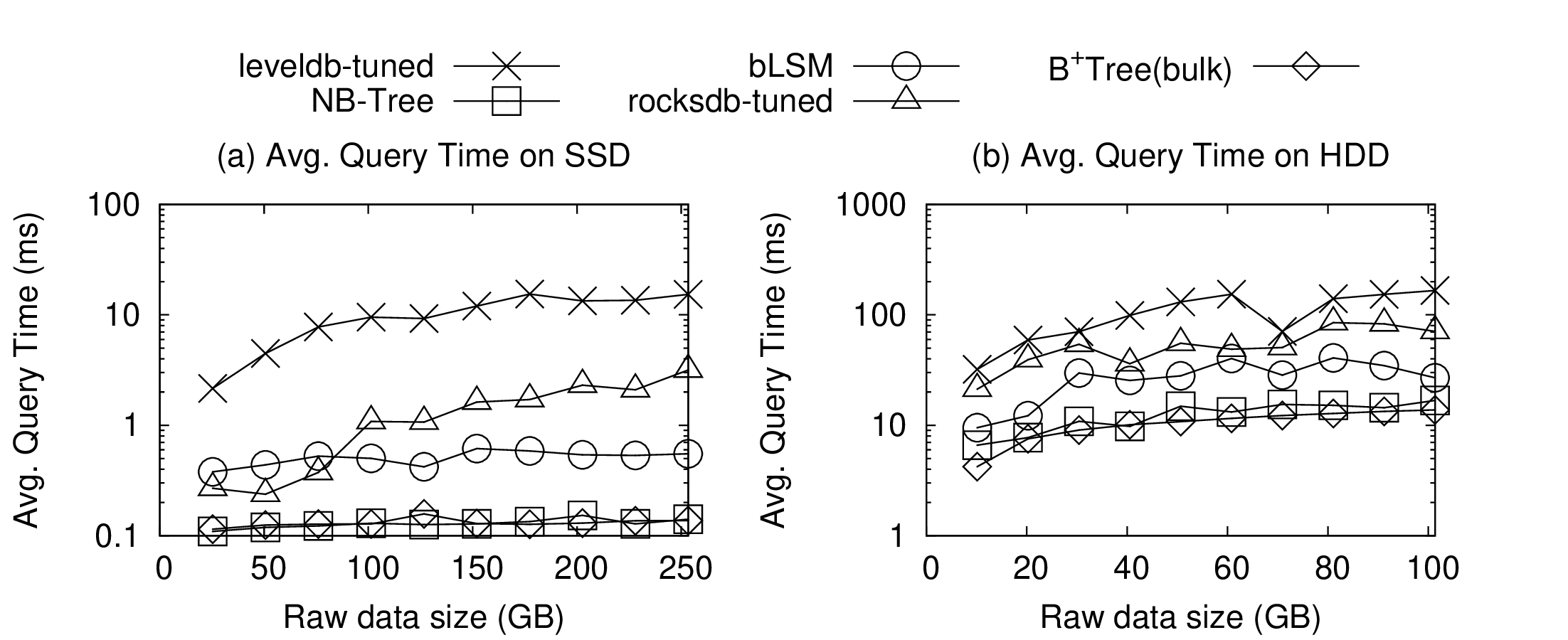}
    	\vspace{-0.7cm}
    	\caption{Avg. Query Time vs. Data Size}
    	\label{fig:exp:query}
    \end{minipage}
    \begin{minipage}[t]{.49\textwidth}
        \centering
    	\includegraphics[width=\textwidth]{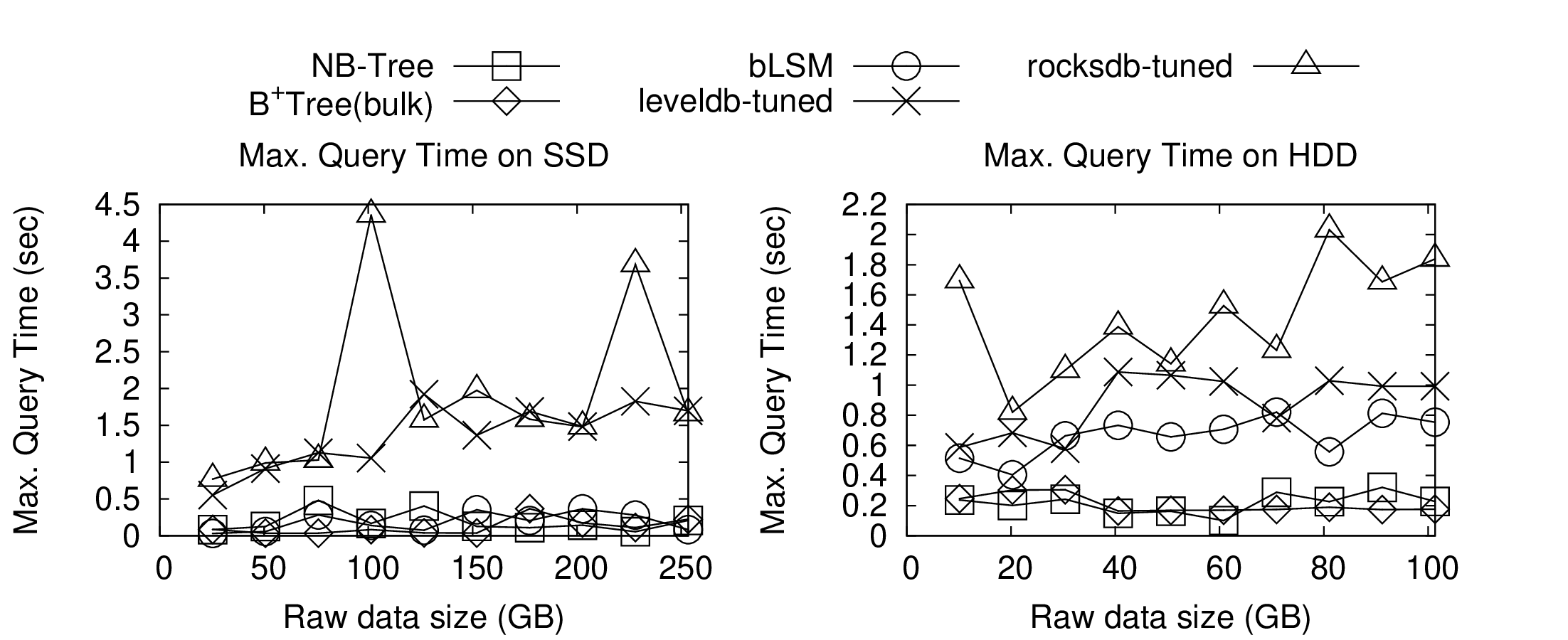}
    	\vspace{-0.7cm}
    	\caption{Max. Query Time vs. Data Size}
    	\label{fig:exp:maxquery}
    \end{minipage}
    \vspace{-6pt}
\end{figure*}

\if 0
\begin{figure*}[t]
	\includegraphics[width=\textwidth]{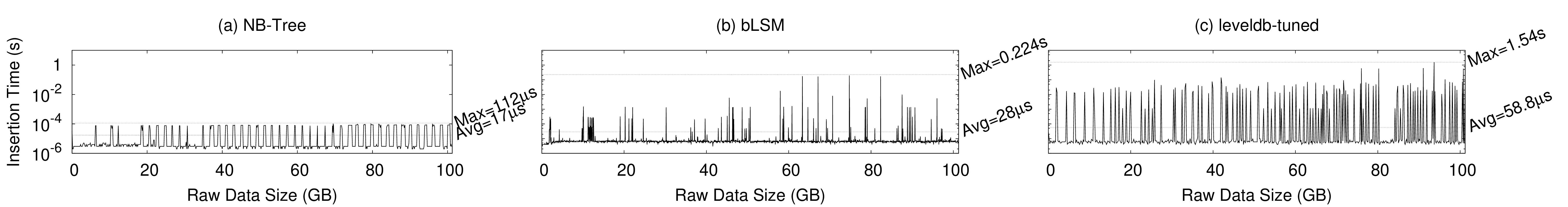}
	\caption{Insertion Time vs. Data Size}
	\label{fig:exp:instInsertionRate}
\end{figure*}
\fi

\vspace{-0.1cm}
\section{Empirical Studies}\label{sec:exp}
\vspace{-0.05cm}
\subsection{Experimental Setup}\label{exp:setup}
We ran our experiment on a machine with Intel Core i5 3.20GHz CPUs and 8 GB RAM running CentOS 7. This machine has (1) a 250GB and 7200 rpm hard disk and (2) an SSD with the model ``Crucial MX500'' and the storage size of 1TB. Each disk page is 4KB.
All algorithms were implemented in C/C++.

\textbf{Dataset.} Following \cite{DI19, DI18, DAM+17}, we conducted experiments on synthetic datasets. Specifically, we generated synthetic datasets with $n$ key-value pairs
where each key is 8 bytes and each value is 128 bytes.
Following \cite{DI19, DI18, DAM+17}, we generated keys uniformly to focus on worst-case performance. 
The largest dataset generated is of size about 250 GB ($2\times 10^{9}$ keys). 

\textbf{Workload.} We designed an \textit{insert} workload and a \textit{query} workload to study the query and insertion performance of different indices. Each \textit{insert} workload is a workload which starts from an empty dataset and involves $n_I$ insertion operations. Each \textit{query} workload is a workload which involves $n_Q$ query operations performed on an index built based on the dataset containing $n$ keys. $n_Q$ is set to $10^4$ throughout the experiments. 
In the query workload, %there are two types of queries, namely \emph{point queries} and \emph{range queries}. For each point query, 
we select keys uniformly from existing keys as the query input. %For each range query, we  randomly select
%two keys as two query input parameters, namely $k_1$ and $k_2$, where $k_1 < k_2$. 
%

%We performed a total of $10^9$, $2\times 10^9$, $10\times 10^9$, $20\times 10^9$ and $50\times 10^9$ operations with the following workloads: (0\% updates, 100\% insertions), (20\% updates, 80\% insertions), (40\% updates, 60\% insertions), (60\% updates, 40\% insertions) and (80\% updates, 20\% insertions). We measure query performance at regular intervals during these workloads. The workloads don't directly contain query because query performance is much slower. As is standard in the literature \cite{DAM+17, KQJ+17, YWH+17,SR12}, the workloads did not include deletions.

\textbf{Measurements.} Based on the four performance metrics discussed in Section \ref{sec:into}, we designed measurements on the indices for each of the two workloads. Consider an \textit{insert} workload involving $n_I$ insertion operations.
We have 2 measurements, namely
(1) \textit{average insertion time} and
(2) \textit{maximum insertion time}.
(1) \textit{Average insertion time} is defined to be
the average time taken per key to finish the entire insert workload, i.e.,  $\frac{t_I}{n_I}$,
where $t_I$ is the total time taken to complete $n_I$ insertion operations. \textit{Average insertion time} helps us verify our theoretical results on \textit{amortized insertion time}.
(2) \textit{Maximum insertion time} is a measure on the entire workload. It is the maximum insertion time of a key over the entire workload. \textit{Maximum insertion time} helps us verify our theoretical results on \textit{worst-case insertion time}.
%
%Note that following \cite{DI19, DI18, DAM+17, SR12}, we adopted ``insertion throughput'' as a measurement instead of ``insertion time'' in this experimental study (since a large throughput means a small query time and vice versa).

Consider a \textit{query} workload involving $n_Q$ query operations.
We have 2 measurements, (1) the \textit{average query time} 
and 
(2) the \textit{maximum query time}. 
(1) The \textit{average query time}  is a measure on the \emph{entire} workload. 
It is defined as the average time taken per key to finish the entire query workload, i.e., $\frac{t_Q}{n_Q}$
where $t_Q$ is the total time taken to complete $n_Q$ query operations in this workload. The \textit{average query time} helps us verify our theoretical results on \textit{average query time}.
(2) The \textit{maximum query time} is a measure on the \emph{entire} workload. 
It is defined to be the maximum query time of a query in the entire workload. 
The \textit{maximum query time} helps us verify our theoretical results on \textit{worst-case query time}.

\if 0

\textbf{Experiments.} To compare the performance of the data structures as more insertions are performed on the data structure, we performed the Insertion workload for various $n_I$ values and measure average and maximum insertion time. Average insertion time relates to amortized insertion time while maximum insertion time relates to worst-case insertion time. To measure the query performance, for a dataset after $n$ operations, we measured average and maximum query time for point queries with uniformly distributed keys. %The length of the range queries were fixed so that the expected number of matching keys is 1000. For expected number of matching keys $c$, the length of the range is calculated by dividing the total range of the keys by the number of keys in the data structure multiplied by $c$ (because distribution of the keys is uniform).

%(based on the definition above) using both uniform and Zipfian query distributions. In a setup similar to that of \cite{DAM+17, KQJ+17, SR12}, we used YCSB \cite{CST+10} to synthetically generate the workload for the Zipfian query distribution (with YCSB's default parameter settings). We also measured the query performance for uniformly distributed range queries of fixed lengths. The length of the range queries were fixed so that the expected number of matching keys is 10, 100, 10000 and 10000. For expected number of matching keys $c$, the length of the range is calculated by dividing the total range of the keys by the number of keys in the data structure multiplied by $c$ (because distribution of the keys is uniform). 

\fi

\textbf{Algorithms.} We compared our index, NB-trees, 
with 6 other indices: (1) LevelDB \cite{G17},
(2) Rocksdb \cite{F17c,F17},
(3) bLSM \cite{SR12}, (4) B$^\epsilon$-tree \cite{BF03} and  (5)
B-tree \cite{BM72} and (6) B$^+$-tree \cite{SKS+97}.  
The first three indices (i.e., LevelDB, Rocksdb and bLSM) 
are three different implementations of LSM-trees. 
Note that there exist many other variants
of the LSM-trees 
 \cite{DI18, LAK16, YHL+17} 
which optimize the insertion/query performance which will be discussed 
in detail in Section~\ref{sec:relatedWork}. 
However, as to be discussed in Section~\ref{sec:relatedWork}, 
these performance optimization techniques originally designed for LSM-trees could also be applied to NB-trees. Thus, these techniques are orthogonal to our work. For fairness, we do not include the other variants of the LSM-trees for comparison. 

Moreover, we ran a preliminary experiment in which we inserted about 6GB of raw data and measured average insertion time of all the algorithms. If average insertion time was larger than $100\mu s$, we excluded the algorithm from the rest of the experiments. This is because based on this result, we can conclude that the algorithm is not suitable for insertion-intensive workload and it will be infeasible to run such an algorithm on the large datasets in our experiment. 

\textit{(1) LevelDB:} LevelDB \cite{G17} is a widely used key-value store
implementing an LSM-tree and has been used in the experiments of many existing studies \cite{DAM+17, SR12, LAK16, WXS+15, SSM+13}. 
In order to have a fair comparison, we adopt two different parameter settings for LevelDB, namely \texttt{leveldb-default} and \texttt{leveldb-tuned}.
\texttt{leveldb-default} is LevelDB with the default setting
similar to \cite{DAM+17, SR12} (i.e., multiplying factor = 10, in-memory write buffer size = 4 MB and no Bloom Filter feature enabled). 
In our preliminary experimental result, we found that the average insertion time
of \texttt{leveldb-default} is larger than $100\mu s$.
In the later experiments, we exclude this algorithm from our experimental results since it could not handle insertions with average insertion time smaller than $100\mu s$.
\texttt{leveldb-tuned} is LevelDB with the ``tuned'' setting for the best-insertion performance. Specifically, in \texttt{leveldb-tuned},
following \cite{DI18, SR12}, we enabled the Bloom Filter feature
using 10 bits per key for short query time.
Due to the large available memory, we varied the user parameter called ``in-memory write buffer size'' from 10 MB and 100 MB to determine the ``best'' buffer size which could give the smallest average insertion time. When the buffer size is larger, LevelDB has fewer merge operations resulting in a smaller insertion time but at the same time, each merge takes longer resulting in a larger insertion time. In our experiment, we found that 32 MB as the ``best'' buffer size. 
Thus, \texttt{leveldb-tuned} is LevelDB with the setting where 
 multiplying factor = 10,  in-memory write buffer size = 32 MB and Bloom Filter feature enabled. 
%Secondly, we experimented with the size of the in-memory write buffer with.
%Both variations use Bloom filters with 10 bits per key. 

\textit{(2) Rocksdb:} Rocksdb \cite{F17c} is a fork of LevelDB with some new features that are not necessarily relevant to our work (e.g., parallelism, see \cite{F17} for details). However, we observed that they performed differently under our workloads, so we include both algorithms. Similar to LevelDB, we performed parameter tuning for Rocksdb and observed that setting the write buffer size to 2GB has the best average insertion time. We refer to this algorithm as \texttt{rocksdb-tuned}. Bloom Filters are enabled and set to 10 bits per key.

\textit{(3) bLSM:} 
bLSM \cite{SR12} is a variant of an LSM-tree proposed for high query performance and low insertion delay. 
For a fair comparison, we obtained a parameter setting of bLSM with the best performance. We varied the user parameter of the in-memory component size to determine the ``best'' in-memory component size with the ``best'' insertion and query performance (increasing memory size improves both insertion and query performance). We found that 6 GB is the ``best'' size. In our experiment, we adopted this setting.

\textit{(4) B$^\epsilon$-trees: }
We implemented two versions of the B$^\epsilon$-trees, namely (a) \emph{Public-Version} 
and (b) \emph{Own-Version}.
(a) \emph{Public-version} is a publicly available version
%called \emph{wiredtiger} which is an implementation of 
of the B$^\epsilon$-trees 
%(and \textit{B$^+$-trees})
used in system TokuDB \cite{P17}. 
%called \emph{wiredtiger} which is an implementation of the B$^\epsilon$-trees 
We adopted the default settings of TokuDB. %\emph{wiredtiger}. 
However, TokuDB's average insertion time in our preliminary experiments is more than $200\mu s$. It was not feasible to run TokuDB in our experiments which requires the insertion time to be at most $100\mu s$.
(b) \emph{Own-Version} is our own implementation of 
B$^\epsilon$-tree.
%(and \textit{B$^+$-trees}). 
\emph{Own-Version} could not handle the insertions with average insertion time less than $100\mu s$.
Thus, since
$B^\epsilon$-tree (both \emph{Public-Version}  and \emph{Own-Version}) is not suitable for high-insertion rate  workloads, we exclude it from our experimental results.

\textit{(5) B-trees and (6) \textit{B$^+$-trees}:}
Similar to B$^\epsilon$-trees, 
we implemented two versions of \textit{B$^+$-trees}, namely \emph{Public-Version} and \emph{Own-Version}.
Here, \emph{Public-Version} denotes the B$^+$-trees used in \emph{wiredtiger} which is a storage engine in MongoDB \cite{W19}.
Similarly, 
we exclude \textit{B-trees} and \textit{B$^+$-trees} in our experimental results
since they could not handle insertions with insertion time smaller than $100\mu s$ per insertion.
However, since it is well-known that B$^+$-trees are good for fast queries,
we implemented a ``bulk-load'' version of a B$^+$-tree 
called \texttt{B$^+$-tree(bulk)}
as a baseline to compare the query performance among all indices in the experiments. We implemented \texttt{B$^+$-tree(bulk)} by 
pre-sorting the data and 
adopting
 a bottom-up bulk-loading approach \cite{SKS+97}. 
We do not include any measurement about the insertion statistics
for \texttt{B$^+$-tree(bulk)}
 since it does not show the realistic insertion performance for \texttt{B$^+$-tree}. 
 The query performance of
  the bulk-load version of a B$^+$-tree (i.e., \texttt{B$^+$-tree(bulk)})
 is better than the ``normal insertion'' version of a B$^+$-tree
 because \texttt{B$^+$-tree(bulk)} could be constructed 
 such that almost all nodes in \texttt{B$^+$-tree(bulk)} are full
 and thus, the data are not scattered across different disk pages,
 resulting in a lower seek time and a smaller query time. 
It is not easy to design a ``bulk load'' version of B-trees
(since some key-value pairs are stored in internal nodes and some are stored in leaf nodes)
and thus, we do not include it.
%Note that by pre-sorting the data for a B$^+$-tree, we constructed a B$^+$-tree with the optimal height, where almost all the nodes are full and not half-empty, and that the nodes do not become scattered around the hard disk (which allows for a lower seek time). Thus, the query performance of the B$^+$-tree shown here will typically be better than B$^+$-trees used in practice. 

%We also included our implementation of a {B$^+$-tree} to evaluate our query performance, referred to as \texttt{B$^+$-tree}.

\textit{NB-Trees.} We implemented the final version of NB-tree discussed in Section \ref{sec:advanced}, referred to as \texttt{NB-Tree}. %we implemented the basic NB-tree discussed in Section \ref{sec:NBTreeDesign} referred to as \texttt{NB-Tree-basic}, and the advanced NB-tree discussed in Section \ref{sec:advanced}, referred to as \texttt{NB-Tree-advanced}. %, which both utilize only one hard disk. To see the impact of using more hard disks on NB-trees, we also implemented NB-tree with M\&S operation that uses three hard disks, referred to as \texttt{NB-Tree-M\&S3D}. 
%We also implemented both versions with Bloom filters (as discussed in Seciton \ref{sec:bloom}), referred to as \texttt{NB-Tree-basic-Bloom} and \texttt{NB-Tree-advanced-Bloom}. 
We set $f$ to 3 and $\sigma$ to 2 GB 
after conducting experiments to find the ``best'' parameter for the NB-tree to be shown in Section~\ref{exp:small:param}.

\vspace{-5pt}
\subsection{Experiment for Parameter Setting}

\begin{table*}
	\centering
	\begin{tabular}{|c|c|c|c|c|c|}\hline
		\multirow{2}{*}{Algorithms}	  &\multicolumn{2}{c|}{\makecell{Amortized insertion time\\ $O(\alpha\times T_{seq, W}+\beta \times T_{seek})$}} 								     &  \multicolumn{2}{c|}{\makecell{Worst-case Insertion Time \\$O(\alpha\times T_{seq, W}+\beta \times T_{seek})$}}	  &\makecell{Worst-case Query Time \\$O(\alpha\times (T_{seq, R}+T_{seek}))$}\\\cline{2-6}
		& $\alpha$ & $\beta$& $\alpha$ & $\beta$& $\alpha$ \\\hline
        B-tree \cite{BM72} &$\log_Bn$&$\log_Bn$&$\log_Bn$&$\log_Bn$ &$\log_Bn$			   \\\hline
		B$^\epsilon$-tree \cite{BF03} &$\frac{f\log_fB}{B}\log_B n$&$\frac{f\log_fB}{B}\log_B n$&			  $\frac{f\log_fB}{B}\log_B n$&$\frac{f\log_fB}{B}\log_B n$ &$\log_fB\log_Bn$   \\\hline
		LSM-tree \cite{OCG+96}	  &$\frac{f\log_fB}{B}\log_B n$&$1$	&$\frac{n}{B}$&$\log_fB\log_Bn$&$\log_fB(\log_B n)^2$\\\hline
		NB-tree (our paper)	  &$\frac{f\log_fB}{B}\log_B n$&$\frac{f\log_fB}{\sigma}\log_B n$ &$\frac{f\log_fB}{B}\log_Bn$&$\frac{f\log_fB}{\sigma}\log_B n$	  &$\log_f\sigma\log_B n$   \\\hline

	\end{tabular}
	\vspace{-0.2cm}
    \caption{Summary of the theoretical results (performance in terms of time)}
	\label{tab:summaryResults}
	\vspace{-0.6cm}
\end{table*}

\vspace{-0.1cm}
%\subsubsection{Parameter Setting}
\label{exp:small:param} 
In this section, our experiment measures the average insertion time for 25GB of raw data ($n_I=2\times 10^8$ keys), and the average query time on a database of size 25GB ($2\times 10^8$ keys). We ran each experiment on an HDD three times and averaged the results, shown in Figs. \ref{fig:exp:paramF}-\ref{fig:exp:paramS}. 

\textit{Fanout.} We studied the effect of fanout $f$ for a small $\sigma$ value, 64MB, and a large $\sigma$ value, 2048MB, on NB-trees. Fig. \ref{fig:exp:paramF} (a) shows that when $\sigma = 64$, increasing $f$ causes average query time to decrease. However, the trend is the opposite when $\sigma = 2048$. This is because query time depends on the number of page accesses and the seek time for the accesses.  When $\sigma$ is small, increasing $f$ reduces the height by a lot (from 8 levels when $f=3$ to 4 levels when $f=15$). When the height is smaller, fewer Bloom filters are checked, decreasing the probability that at least one of the Bloom filters returns a false positive. Thus, increasing $f$ reduces the number of page accesses and the query time. However, for large values of $\sigma$, increasing $f$ does not change the height by much (from 4 levels when $f=3$ to 3 levels when $f=15$). In this case, most queries perform only one disk access. Note that, d-trees of sibling s-nodes are written sequentially to the disk. Thus, when $f$ is large, keys that are close to each other in the key space are written close to each other on disk. However, the query distribution is uniform, and it is likely that consecutive query keys are not close to each other in the key space. Hence, when $f$ is large, the seek time during the queries becomes larger. This is less of an issue when $f$ is small. Therefore, increasing $f$ increases the seek time for queries. As a result, for $\sigma=2048MB$, query time worsens when $f$ increases.

Fig. \ref{fig:exp:paramF} (b) shows that the insertion time increases when $f$ increases. This result generally follows the theoretical model where the factor $f$ in amortized insertion time complexity causes the insertion time to increase when $f$ gets larger.

\textit{D-tree size.} Fig. \ref{fig:exp:paramS} shows that, generally, larger $\sigma$ improves insertion time but worsens the query time, as theory suggests. However, one interesting observation is a local minimum observed at $\sigma = 16MB$ for average insertion time in Fig. \ref{fig:exp:paramS} (b). This can be attributed to the HDD cache being $16MB$, which improves the sequential I/O performance during $HandleFullSNode$. As $\sigma$ gets beyond $4GB$, the insertion time increases since \texttt{NB-Tree} does not fit in main memory. The improvement in query performance when $\sigma$ is larger than $1GB$ is because the main memory component becomes large compared to data size and some of the queries are answered by just checking the in-memory component. 

\textit{Parameter Setting.} In the rest of the experiments, we optimize NB-tree for an insertion-intensive workload. We select $\sigma=2GB$ which has the best insertion performance based on Fig. \ref{fig:exp:paramS} (b) and set $f=3$ because for $\sigma=2GB$, in Fig. \ref{fig:exp:paramF} (b), $f=3$ has the best insertion performance. Based on this parameter setting, we note that \texttt{NB-Tree}'s memory usage is as follows. For data size of 250GB (the maximum data size used in our experiments), about 2.3GB is allocated for caching Bloom filters and 1GB for caching non-leaf node of d-trees. Interestingly, even when optimizing NB-trees for insertions, they perform queries almost as fast as a B$^+$-tree. 

\vspace{-5pt}
\subsection{Experiment for Baseline Comparison}

\textit{Average insertion time.} Fig. \ref{fig:exp:insertion} shows the average insertion time of the indices on HDD and SSD. \texttt{NB-Tree} achieves the lowest time on both HDD and SSD, while \texttt{bLSM}'s performance deteriorates when the data size gets larger because it keeps the number of components constant. \texttt{rocksdb-tuned} performs similar to \texttt{NB-Tree} on HDDs but its performance is worse on \texttt{SSD}. Note that the performance advantage of \texttt{NB-tree} compared with \texttt{rocksdb-tuned} and \texttt{bLSM} is more visible on SSDs. This shows that NB-trees perform better on larger data sizes when the ratio between data size and in-memory component is larger. %We expected \texttt{leveldb-tuned} to have the average insertion time similar to NB-tree, which could not be shown in the figure. We suspect that the worse performance of \texttt{leveldb-tuned} is due to system overheads, such as concurrency control or logging, or parameter setting issues, as mentioned in \cite{SR12}. 

%Regarding \texttt{rocksdb-tuned}, fluctuations in performance can be observed, more significant on HDD. We ran the algorithm several times to ensure the fluctuations were not a result of randomness and observed similar results in these instances. Note that for an LSM-tree, in practice, average insertion time can depend on how full the levels close to the root are, since it impacts the time merging the root takes. For \texttt{rocksdb-tuned}, the size of the write-buffer is large (compared with \texttt{leveldb-tuned}), resulting in larger fluctuations in size the of the levels, as different levels are merged. Thus, average insertion time fluctuates more compared with \texttt{leveldb-tuned} (recall that average insertion time is measure over). However, \textt{NB-tree}'s structure does not significantly change over time, resulting in a stable performance. 

\textit{Maximum insertion time.} Fig \ref{fig:exp:maxinsertion} shows the maximum insertion time of the indices. \texttt{NB-Tree} achieves the lowest time on both HDD and SSD, outperforming other algorithms by at least 1000 times for some data sizes on both HDD and SSD. Maximum insertion time of \texttt{rocksdb-tuned}, \texttt{bLSM} and \texttt{leveldb-tuned} goes as high more than 0.2 s (for \texttt{rocksdb-tuned}, this number is 453s), which is unacceptable for many applications. The superior performance of \texttt{NB-Tree} is due to their logarithmic worst-case time together with the deamortization mechanism suggested in Section \ref{sec:advanced}. Observe that \texttt{rocksdb-tuned} has the maximum insertion time of 453 seconds. Even though that happens only once during the insertion processes, it makes the system unreliable. %In these experiments we measure the insertion rate at each point of time when performing the insertion workload. Our goal is to see how low the insertion throughput drops for each of the data structures, and how frequently that happens. Fig \ref{fig:exp:instInsertionRate} shows the result of this experiment. For clarity, we have plotted every $10^6$-th value from the original data points for bLSM and NB-tree and every $2*10^6$-th value for leveldb-tuned  (but we included the 4 lowest instantaneous throughput values for all algorithms before the sampling). 

%First, note that \texttt{NB-Tree}'s minimum throughput is more than 200 times higher than \texttt{leveldb} and \texttt{bLSM}. Furthermore, for both \texttt{bLSM}'s and \texttt{leveldb}'s throughput drops to below 10 ins/per second which is unacceptable.  

\textit{Average query time.} Fig. \ref{fig:exp:query} shows the average query time of the indices. \texttt{NB-Tree} achieves query time almost as low as \texttt{B$^+$-tree(bulk)} (which is worst-case optimal). \texttt{rocksdb-tuned}, \texttt{leveldb-tuned} and \texttt{blsm} have query times larger than \texttt{NB-Tree}, more prominently on SSDs.
%Here, we discuss the performance for queries with uniform distribution. %Results for Zipfian distribution are similar and for the sake of space are shown in 
%\ifx\techReport\undefined
%our technical report \cite{techreport}.
%\else
%Section \ref{sec:appndx:exp}.
%\fi

\textit{Maximum query time.} Fig. \ref{fig:exp:maxquery} shows the maximum query time of the indices. \texttt{rocksdb-tuned} has the worst performance while \texttt{B$^+$-tree(bulk)} is generally better. Note that all queries have to wait for at least one disk I/O operation, but an I/O operation can take long if the operating system is busy or if there are disk failures. Thus, maximum query time has a large variance and the comparison among the algorithms is less conclusive (note that insertions do not need to wait for disk I/O operations due to in-memory buffering). 

%\textit{Without Bloom filters.} Based on Fig. \ref{fig:exp:query} (a), query performance of \texttt{NB-Tree-basic} and \texttt{NB-Tree-advanced} is much better than \texttt{leveldb-tuned}, and their increase in query time is much smaller when the size of the dataset increases, depicting the logarithmic factor difference between the query time of LSM-trees compared with an NB-Tree and a B$^+$-tree. With a small constant factor difference, \texttt{B$^+$-tree} performs better than \texttt{NB-Tree-basic} and \texttt{NB-Tree-advanced}. 

%\textit{With Bloom filters.} As Figure \ref{fig:exp:query} (b) shows, adding Bloom filters to the algorithms results in a performance improvement for all of algorithms, although the performance of \texttt{leveldb-bloom} still significantly lags behind the others. NB-tree data structures with Bloom filters, achieve a similar query performance as \texttt{B$^+$-tree} for large datasets. 

\textbf{Summary.} 
The average query time of an NB-tree is 4 times smaller than LevelDB
and 1.5 times than bLSM and Rocksdb. 
%An NB-tree provides its query time about 4 times faster than LevelDB and 1.5 times faster than bLSM.
It is similar to the average query time of 
a nearly optimally constructed, bulk-loaded B$^+$-tree, 
where building a B$^+$-tree incrementally takes orders of magnitude longer than an NB-tree. 
Besides, the average and maximum insertion time of an NB-tree
(which are at most 0.0001s)
are multiple factors smaller than LevelDB, Rocksdb and bLSM
(which could be greater than 0.2s).
Overall, an NB-tree provides a more reliable insertion and query performance. 

%Finally, it is worth mentioning that \texttt{leveldb-tuned} at insertion rate 100,000 ins./sec (in Figure \ref{fig:exp:insertion} (c), for $N= 10^{10}$, total delayed insertion time is about 24,958 seconds) has, in fact, larger total delayed insertion time compared with NB-Tree data structures at insertion rate 200,000 (in Figure \ref{fig:exp:insertion} (d),  at $N =10^{10}$ \texttt{NB-Tree-BM\&S} and \texttt{NB-Tree-M\&S} have total delayed insertion time of 3711 and 16658 seconds). This shows empirically that the NB-Tree data structure is better at performing consistently at a certain insertion rate, confirming the theoretical result on the insertion rate guarantee of \texttt{NB-Tree-BM\&S}, while \texttt{leveldb-tuned}'s insertion performance can vary largely based on the state of the data structure. 

%of various data structures with different features. Previously mass-tree \cite{MKM12} has been proposed as an in-memory nested structure where the structural tree is a trie.  
%TODO: cite the other works with nested structure, mention differences. I think the other ones have only used hashing for nesting.

\vspace{-0.3cm}
\section{Related Work}\label{sec:relatedWork}
We discuss indices used for insertion intensive workloads. 

\textbf{LSM-trees.} LSM-tree is an index used for insertion-intensive workloads used in many systems such as BigTable \cite{CDG08}, LevelDB\cite{G17}, Cassandra \cite{LM10}, HBase \cite{ABC+12}, RocksDB \cite{F17}, Walnut \cite{CDM+12} and Astrix DB \cite{ABB+14}. By using an in-memory component and several on-disk B-tree components, LSM-trees \cite{OCG+96} perform very few seek operations during insertions. However, this design causes a sub-optimal number of I/O operations during queries, and linear worst-case insertion time that causes long insertions delay (see \cite{TY17, LHY+10} for a discussion of LSM-tree's performance). Many improvements have been proposed to LSM-trees' design as discussed below.

\textbf{Query improvement.} \cite{SR12} uses Bloom filters to improve the query time and \cite{DAM+17} tunes the Bloom filter parameters. Compared with LSM-trees, we showed that Bloom filters adopted by NB-trees provide better theoretical and empirical performance. Method of \cite{DAM+17} can also be used by NB-trees to optimize the Bloom filter parameters. Moreover, \cite{JOE+07, DFF17} partition an LSM-tree into several smaller LSM-tree components which provides a constant factor improvement. 

\cite{LHY+10} uses fractional cascading \cite{BG86} to provide asymptotically optimal worst-case query time. Fractional cascading connects different LSM-tree components to each other. Consider the B$^+$-tree of the $i$-th level of the LSM-tree. In each leaf node, $N$, of the B$^+$-tree, some key-value pairs have extra pointers pointing to a node, $N'$, of the $(i+1)$-th level. The pointers from the $i$-th level to the $(i+1)$-th level are called \textit{fence pointers}. Fence pointers satisfy the properties that (1) the first key-value pair $k$ of node $N$ must have a fence pointer pointing to a next-level node $N'$ and every node $N'$ at level $i$+1 must have a fence pointer pointing to it from level $i$. %Since there are more nodes at level $i+1$ than level $i$, nodes in level $i$ may have more than one fence pointer to different nodes in the $(i+1)$-th level (they have at least one because the first key-value pair always has a fence pointer).
(2) Consider two keys, $k_s$ and $k_l$, in level $i$ that have fence pointers to nodes $N_s$ and $N_l$ in level $i+1$, such that there does not exist another key $k$ in level $i$ that has a fence pointer and that $k_s <k<k_l$. Let $r_s$ be the smallest key in $N_s$ and $r_l$ the smallest key in $N_l$. It holds that $r_s \leq k_s < k_l \leq r_l$. These properties help in performing a constant number of disk-page accesses at each level.

LSM-trees with fractional cascading suffer from large worst-case insertion time and are not compatible with Bloom filters \cite{SR12}. Thus, they provide a worse query performance in practice. The reason for their incompatibility is that to search the $(i+1)$-th level using the $i$-th level fence pointers, we need to have searched the $i$-th level.
Based on this deduction, we need to have searched all the levels of the LSM-tree. However, using Bloom filter is only advantageous when we do not need to search all levels of the LSM-tree.

\textbf{Insertion improvement.} Most of the focus has been on optimizing the merge operation, divided into \textit{leveling} and \textit{tiering} categories. \textit{leveling} is the category discussed so far, which  sorts each LSM-tree component during the merge. \textit{Tiering}, during a merge operation, appends the data to the lower levels and only sorts a level after it is full. This avoids rewriting the lower level component during the merge operation at the expense of the query time. \cite{DI18} uses the leveling merge policy at some levels of the tree and tiering merge policy at other levels. In \cite{DI19} unlike the original design, the ratio of the size across different adjacent levels of the LSM-tree is not constant. More variations of tiering are discussed in \cite{ZXL+16, BDG+17, YWH+17, WXS+15, PYX17, YHL+17}. \cite{BBH+18} discusses in-memory optimization for faster writes. These improvements are orthogonal to our work and can be adopted by NB-trees in the future. \cite{LAK16} discusses a theoretical model to analyze insertion performance of LevelDB and provides methods for parameter optimization. Their methods require knowledge of probability distribution of the keys in advance and performs time-consuming optimizations not feasible in the real-world. Thus we did not include their method in our experiments. \cite{TY17, LHY+10, SR12} discuss reducing the worst-case insertion time, but their methods take linear time to the data size compared with the logarithmic worst-case time of NB-trees.

\textbf{B-tree and B-tree with Buffer.} B-trees \cite{BM72} are read-optimized indices, performing optimal number of I/O operations during queries\cite{BF03}. But they perform a seek operation for every page access, sacrificing their insertion performance. B-trees with Buffer \cite{BF03} (also known as B$^\epsilon$-trees) are a write-optimized variant of B-trees where part of each disk page allocated to each node is reserved for a buffer. The buffer is flushed down the tree when it becomes full. B-trees with Buffer can be seen as a special case of NB-trees where s-node size is one disk page and their analysis of query and insertion performance follows from that of NB-trees. In such a case, all disk accesses involve a seek operation, worsening the insertion performance, as our experiments confirmed. They also have worse space utilization since they allow half full nodes and worse range query performance since their nodes are not written sequentially on the disk. NB-trees keep their d-nodes full and write them sequentially for each s-node.

\textbf{Other data structures.} Many write optimized data structures such as \cite{BFJ+17, IP12, WSJ+14, LPG+17} have been proposed for a variety of settings and we do not have space to cover them all. Among them, Y-tree \cite{JDO99} is similar to B-trees with Buffer but allows for larger unsorted buffers at each non-leaf level of the B-tree that reduces the number of seek operations performed during insertions (can also be seen as a form of \textit{tiering}). For a buffer similar in size to that of B-tree with Buffer, their performance will be similar to B-trees with Buffer and with the same weaknesses. However, a larger buffer worsens the point query performance (although range queries will not be affected as adversely), since it requires searching multiple pages of the unsorted buffer at each level of the tree by long scans. Y-trees also suffer from the issues mentioned above regarding space utilization and seek operations during range queries of B-trees with Buffer. Finally, mass-tree \cite{MKM12} is an in-memory data structure that is similar to this paper using a nested index, but the structural tree for mass-tree is a trie which, although works well in memory, can be unbalanced and cause large insertion and query cost if adopted for secondary storage. 

In-memory optimization is outside the scope of this paper, but in-memory optimizations for B-trees such as \cite{LLS13, RR00} improve the in-memory performance. However, their on-disk insertion performance is the same as B-trees, which is worse than NB-trees in terms of amortized insertion time.   

\textbf{Summary.} Table \ref{tab:summaryResults} shows the theoretical performance of the indices mentioned above (written as multiples of $\log_B n$ for easier comparison). For amortized insertion time, NB-trees perform $\frac{\sigma}{B}$ times fewer seek operations than B$^\epsilon$-trees, $\frac{\sigma}{f\log_fB}$ times fewer than B-trees, and similar to LSM-trees ($\sigma$ is typically in the order of 10,000 times larger than $B$). NB-trees have worst-case insertion time logarithmic in data size while LSM-trees' worst-case insertion time is linear in data size. NB-trees' query time is a factor $\log_{\sigma} n$ smaller than LSM-trees and is asymptotically optimal. Overall, NB-trees have a better worst-case insertion and query time (considering the number of seek operations) than existing indices while maintaining practical properties, such as compatibility with Bloom filters and high space utilization.

\vspace{-0.3cm}
\section{Conclusion}\label{sec:conclusion}
We introduced Nested B-trees, an index that theoretically guarantees logarithmic worst-case insertion time and asymptotically optimal query time, and thus supports insertions at high rates with no delays while performing fast queries. This significantly improves on LSM-trees' linear worst-case insertion time and suboptimal query time and avoids long delays that frequently occur in LSM-trees during insertions. We empirically showed that NB-trees outperform RocksDB \cite{F17c}, LevelDB \cite{G17} and bLSM \cite{SR12}, commonly used LSM-tree databases, performing insertions faster than them and with maximum insertion time of 1000 smaller and lower query time by a factor of at least 1.5. NB-trees perform queries as fast as B-trees on large datasets, while performing insertions at least 10 times faster. In the future, a more detailed study can be done on optimizing in-memory caching of the metadata, optimizing the parameter setting of Bloom filters, and using different flushing schemes such as tiering. 

\clearpage

\bibliographystyle{abbrv}
\bibliography{NestebBTree}

\ifx\techReport\undefined
%our technical report \cite{techreport}
\else
%\appendix
%\input{Appendix}
\fi

\end{sloppy}
\end{document}